\documentclass[conference]{IEEEtran}
\IEEEoverridecommandlockouts

\def\BibTeX{{\rm B\kern-.05em{\sc i\kern-.025em b}\kern-.08em
		T\kern-.1667em\lower.7ex\hbox{E}\kern-.125emX}}

\usepackage[linesnumbered,ruled,vlined]{algorithm2e}

\usepackage{url}
\usepackage{verbatim}
\usepackage{graphicx}
\usepackage{multirow}
\usepackage{balance}
\usepackage{lscape}
\usepackage{float}
\usepackage{xcolor}
\usepackage{paralist}
\usepackage{booktabs}
\usepackage{xspace}
\usepackage{listings}
\usepackage{lipsum}
\usepackage{subcaption}
\usepackage{dsfont}

\usepackage{amsmath}

\definecolor{ForestGreen}{rgb}{0.0, 0.66, 0.47}
\definecolor{RubineRed}{rgb}{1.0, 0.0, 0.31}

\newtheorem{example}{Example}

\newcommand{\name}{\ensuremath{\mathsf{MageSQL}}\xspace}

\begin{document}

\title{MageSQL: Enhancing In-context Learning for Text-to-SQL Applications with\\ Large Language Models}

\author{
\IEEEauthorblockN{Chen Shen}
\IEEEauthorblockA{\textit{Megagon Labs}\\
	chen\_s@megagon.ai
}
\and 
\IEEEauthorblockN{Jin Wang}
\IEEEauthorblockA{\textit{Megagon Labs}\\
	jin@megagon.ai
}
\and
\IEEEauthorblockN{Sajjadur Rahman}
\IEEEauthorblockA{\textit{Megagon Labs}\\
  sajjadur@megagon.ai
}
\and
\IEEEauthorblockN{Eser Kandogan}
\IEEEauthorblockA{\textit{Megagon Labs}\\
  eser@megagon.ai
}
}

\maketitle

\begin{abstract}
 The text-to-SQL problem aims to translate natural language questions into SQL statements to ease the interaction between database systems and end users.
 Recently, Large Language Models (LLMs) have exhibited impressive capabilities in a variety of tasks, including text-to-SQL.
 While prior works have explored various strategies for prompting LLMs to generate SQL statements, they still fall short of fully harnessing the power of LLM due to the lack of (1) high-quality contextual information when constructing the prompts and (2) robust feedback mechanisms to correct translation errors.
 To address these challenges, we propose \name, a text-to-SQL approach based on in-context learning over LLMs. 
 \name explores a suite of techniques that leverage the syntax and semantics of SQL queries to identify relevant few-shot demonstrations as context for prompting LLMs. 
 In particular, we introduce a graph-based demonstration selection method --- the first of its kind in the text-to-SQL problem --- that leverages graph contrastive learning adapted with SQL-specific data augmentation strategies. 
 Furthermore, an error correction module is proposed to detect and fix potential inaccuracies in the generated SQL query.
 We conduct comprehensive evaluations on several benchmarking datasets.
 The results show that our proposed methods outperform state-of-the-art methods by an obvious margin.
\end{abstract}

\section{Introduction}\label{sec-intro}

Given a relational database, the text-to-SQL problem automatically translates the natural language question into an SQL statement that queries the database system to find the results.
This problem is increasingly critical for improving the accessibility and usability of relational database systems for a broad range of users, especially non-technical users ~\cite{DBLP:journals/pvldb/FuLWLTS23,DBLP:journals/pvldb/GaoWLSQDZ24,DBLP:journals/pvldb/KimSHL20} who are not familiar with database concepts and SQL. 
% \saj{adding as a placeholder citation. will try to find something in hci literature.}
% Jin: these claims are general ones from previous work of Text-to-SQL for the motivation, I just add related several ones with similar claims.

There is a long stream of research on the topic of text-to-SQL from both database and NLP communities.
Earlier studies~\cite{DBLP:journals/pvldb/SahaFSMMO16,DBLP:journals/pvldb/KimSHL20} employed rule-based methods that first converted the natural language question into an intermediate representation and then mapped them into SQL abstract syntax trees with heuristic rules.
Later, techniques emerged that utilized deep learning techniques to develop solutions~\cite{DBLP:conf/emnlp/ScholakSB21,DBLP:conf/emnlp/GanCXPWDZ21,DBLP:conf/emnlp/QiTHW0ZWZL22,DBLP:conf/aaai/LiHCQ0HHDSL23}, which can support cross-domain adaption as well as handle complex queries.
The basic idea is to formulate text-to-SQL as a machine translation problem and then utilize different variants of models with encoder-decoder architecture to solve it.
Follow-up work such as ~\cite{DBLP:journals/coling/ChoiSKS21,DBLP:journals/pvldb/FuLWLTS23,DBLP:conf/aaai/Li00023} further proposed sketching-based solutions to regularize the syntax of generated SQL queries via pre-defined templates.

Most recently, advances in the era of Large Language Models (LLMs) have brought new opportunities to the problem of text-to-SQL.
Pre-trained LLMs such as GPT-4~\cite{DBLP:journals/corr/abs-2303-08774}, LLaMA~\cite{DBLP:journals/corr/abs-2302-13971} and Codex~\cite{DBLP:journals/corr/abs-2107-03374} have shown superior abilities in understanding human instructions as well as generating structured output.
Such LLMs are generative models that take a sequence of tokens as input and generate a sequence of tokens as output. 
Various studies have shown that input to the models, i.e., prompts, is critical in achieving desired results. 
As such, \emph{prompt engineering} has become an important methodology in utilizing LLMs~\cite{DBLP:journals/csur/LiuYFJHN23}. 
Consequently, departing from previous algorithmic approaches, the LLM-based solutions focused on engineering effective prompt strategies to improve the overall performance ~\cite{DBLP:conf/nips/PourrezaR23,DBLP:conf/emnlp/Nan0ZRTZCR23,DBLP:journals/pvldb/GaoWLSQDZ24}.

\begin{figure}[ht]
	\centering
	\begin{subfigure}{\columnwidth}
		\centering
		\includegraphics[width=0.9\textwidth]{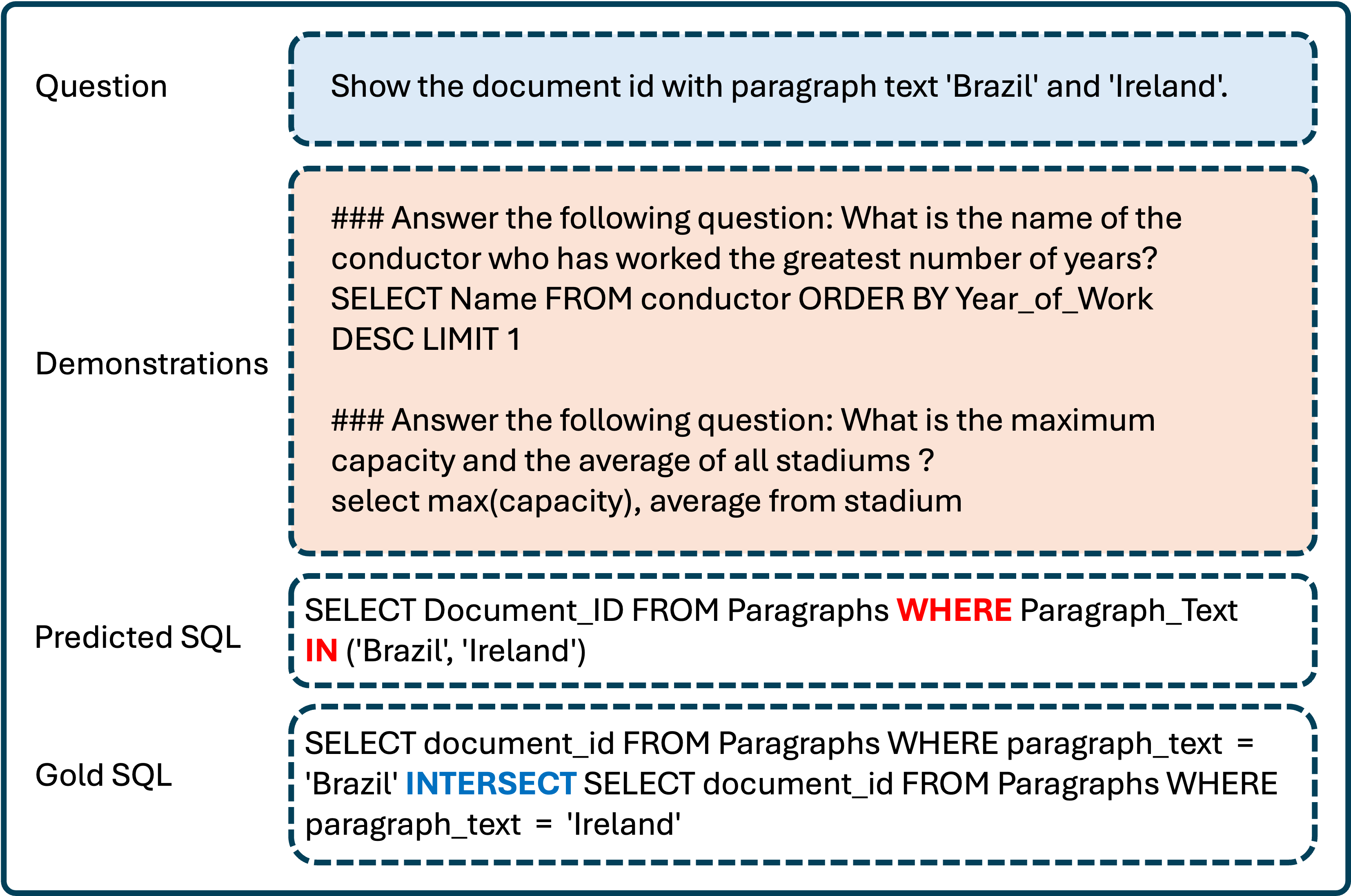}
	\end{subfigure}
	\begin{subfigure}{\columnwidth}
		\centering
		\includegraphics[width=0.9\textwidth]{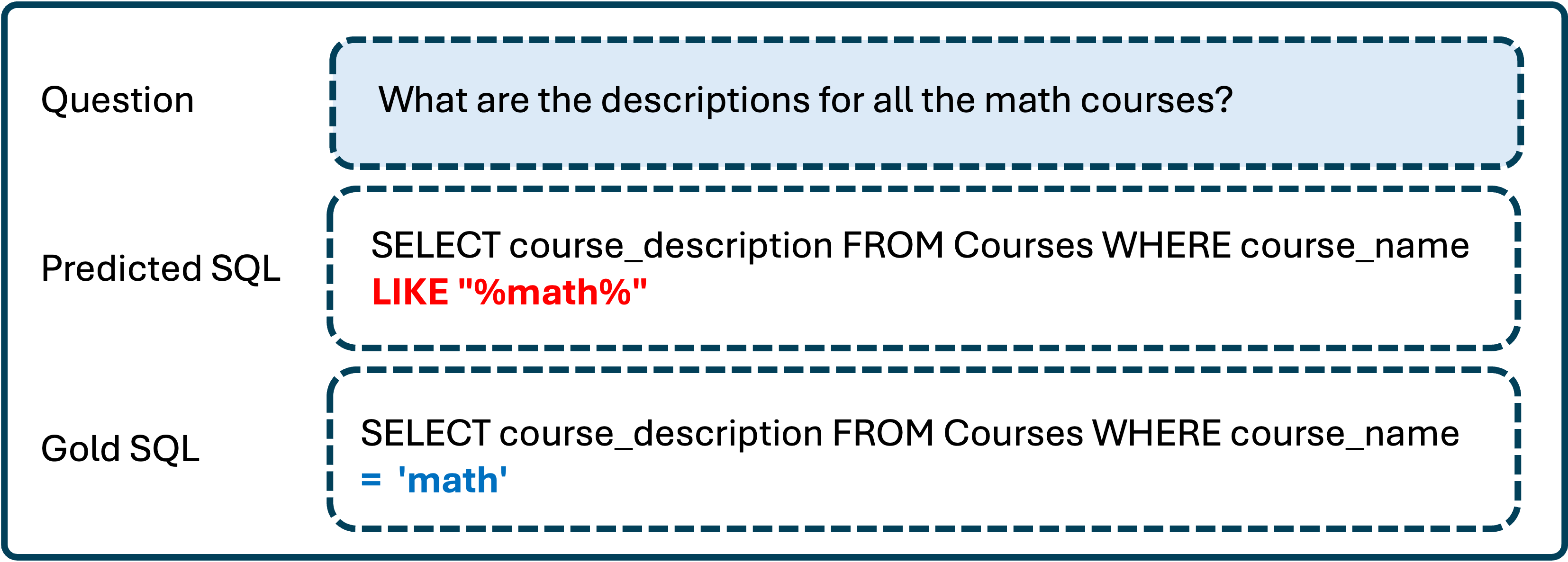}
	\end{subfigure}
	\caption{Motivation Examples}
	\label{fig:motivation}
\end{figure}

However, some research challenges remain unresolved regarding fully utilizing the impressive capabilities of LLMs. 
Firstly, previous results show that while LLM-based solutions are good at understanding natural language questions, there are still various issues in generating SQL statements.
This is mainly due to the lack of effective examples in the prompt that guide the LLM for SQL generation.
Secondly, while previous LLM-based solutions focused on developing advanced reasoning techniques, e.g. chain-of-thought~\cite{DBLP:conf/nips/Wei0SBIXCLZ22}, to facilitate the generation process, they treated the output of LLMs as the final results to be executed in the database systems.
However, since it is well known that the output of LLMs might have uncertainties such as hallucinations, such a practice in previous studies might fail to address the potential issues in the output. 
Let's illustrate these through a few examples in Figure~\ref{fig:motivation}:

\begin{example}
 As shown in Figure~\ref{fig:motivation}, the top example illustrates a case with two-shot learning, where the ground truth requires the usage of conjunctions, yet the demonstration examples did not include any. 
 Consequently, without proper guidance, an LLM may struggle to detect the need for a conjunction, leading to potential mistakes.
 The bottom example shows a case where the prediction from LLM is in fact very close to the ground truth. 
 However, there is a slight mismatch between the LLM output and ground truth, which could be easily fixed by rule-based format adjustment.
 These examples clearly illustrate the need for proper guidance of LLM generation through better demonstration examples and post-processing to resolve final issues, even when the LLM successfully understands the query intent.
\end{example}

% \saj{I feel like we need clarify (potentially in Section 2.2) our assumption of existence of an example pool and justify the why that makes sense. For example, it's expected to have previous logs of user queries-SQL pairs that can be used as example pool.}
% Jin: thanks for the suggestion, add in the second paragraph of Section 3.2 when introducing the techniques for demonstration selection as following: The candidates of demonstration examples could be pairs of natural language question and corresponding SQL that do not appear in the dev/test set.  And mention in Section 4.1: We use the training set as the candidate of demonstration examples for few-shot learning and report the results on dev and test sets.

In this paper, we propose \name, a new framework for Text-to-SQL based on in-context learning over LLMs. 
First, our key observation is that high-quality demonstration examples in few-shot learning is important to improving effectiveness. 
Toward this goal, it is critical that the examples are similar to the questions and potential answers to provide meaningful guidance to SQL generation.
We explore strategies motivated by previous work and recognize that it is crucial to consider both structural and semantic information in the selection to ensure that better examples with a similar structure to the ground-truth SQL are included.
To this end, we first develop a structure-based solution that uses the similarity between Abstract Syntax Trees of SQL statements as the metric for demonstration selection. 
Next, we develop a graph embedding-based solution that can capture both structural similarity and semantics of SQL statements.
This could be realized by constructing a graph for each SQL, which consists of both the syntactic parsing results of SQL and the schema of tables associated with it.
Then the similarity between SQL statements could be evaluated via that between their graph embeddings.
To reach this goal, we propose a graph contrastive learning~\cite{DBLP:conf/iclr/SunHV020,DBLP:conf/nips/0001XLW21,DBLP:conf/kdd/HouLCDYW022,DBLP:conf/nips/YouCSCWS20} based framework to learn the graph encoder for node embedding in a fully unsupervised manner. Therefore, the framework is generalizable to any domain and easier to adopt as no human supervision via annotation is required. 
Then the embedding of an SQL, i.e., the whole graph, could be obtained by aggregating the embedding of all its nodes.

In addition, as part of post-processing, we develop a new error correction module to fix the potential errors in the output.
Specifically, we employ two categories of error-correction strategies: rule-based and prompt-based.
The rule-based strategy aims to correct minor syntax and string format errors and, thus, is very lightweight. 
On the other hand, the prompt-based strategy would do another round of prompts by asking an LLM to rewrite the generated SQL using a set of predefined guidelines so as to correct the errors. 
The goal of this approach is to resolve more complicated errors that cannot be easily handled by hand-crafted rules. 
This is inspired by recent work~\cite{DBLP:journals/corr/abs-2308-08155,DBLP:conf/nips/SchickDDRLHZCS23} of \emph{multi-agent system}, where complex tasks are finished through a collaboration of multiple agents, making our efforts in each step suitable for deployment in such systems as independent agents~\cite{DBLP:conf/cikm/00090RK24}.

The contribution of this paper is summarized as follows:
\begin{itemize}
	\item We propose a new framework to enhance in-context learning for the Text-to-SQL task based on LLMs.
	\item We investigate the strategies of demonstration selection under the few-shot setting and employ structure similarity to find high-quality examples.
        \item We introduce the first-of-its-kind graph embedding-based solution for demonstration selection in text-to-SQL, which resulted in up to 5.4\% performance gain over previous selection methods.
	\item We developed a novel error correction module to fix the potential errors in the generated SQL to improve the overall performance.
	\item We conducted extensive experiments on two popular text-to-SQL benchmarks by up to 13.2\% in Execution Accuracy compared to previous LLM-based solutions. 
	The results showed that our proposed method outperformed state-of-the-art methods.
\end{itemize}

The rest of the paper is organized as follows:
Section~\ref{sec-prelim} provides essential background about LLM and the text-to-SQL problem.
Section~\ref{sec-method} presents our proposed framework.
Section~\ref{sec-exp} introduces the evaluation results.
Section~\ref{sec-related} surveys the related works.
Section~\ref{sec-conc} concludes the whole paper.

\section{Preliminary}\label{sec-prelim}

\subsection{Large Language Model Terminologies} \label{subsec-llm}

Recent years have witnessed a rapid advance in the application of large-scale, pre-trained language models in almost all NLP tasks.
The milestone work of pre-trained Language Model (PLM) is \textsf{BERT}~\cite{DBLP:conf/naacl/DevlinCLT19}  that aimed at learning contextual word embeddings by pre-training a bi-directional transformer-based architecture, comprising a stack of self-attention layers that calculates distributed representations based on the similarity against all tokens and produces contextual embeddings for each input token.
There are two steps in the development of PLMs: pre-training and fine-tuning.
In the pre-training step, the language model is trained on a large unlabeled corpus such as Wikipedia to gain deep language understanding via the pre-training tasks.
The pre-trained model could be further fine-tuned for specific target downstream tasks with labeled training data.

Large Language Models (LLMs) have emerged as a new paradigm of research works in a variety of research fields.
Many pre-trained LLMs have been released to provide public APIs or checkpoints, such as GPT~\cite{DBLP:journals/corr/abs-2303-08774}, LLaMA~\cite{DBLP:journals/corr/abs-2302-13971}, Palm~\cite{DBLP:journals/jmlr/ChowdheryNDBMRBCSGSSTMRBTSPRDHPBAI23} and CodeX~\cite{DBLP:journals/corr/abs-2107-03374}.
Compared with PLMs,  LLMs are pre-trained following similar methodologies but have a much larger number of parameters.
For example, the number of parameters of pre-trained BERT and GPT-3 is 340 million and 178 billion, respectively.
Due to their huge size, a common way to utilize LLM without incurring significant overhead is to provide text instructions to guide generation, known as prompt engineering~\cite{DBLP:journals/csur/LiuYFJHN23}.
LLMs have demonstrated remarkable in-context learning abilities~\cite{DBLP:conf/nips/BrownMRSKDNSSAA20}, guiding predictions based on a relatively few pieces as additional input.
Generally speaking, there are two different settings of in-context learning: (1) \emph{few-shot learning}, when demonstration examples are included in the prompt as input; (2) \emph{zero-shot learning}, when no demonstration is presented.

\subsection{The Text to SQL Problem}\label{subsec-t2s}

\begin{figure}[ht]
	\centering
	\includegraphics[width=0.45\textwidth]{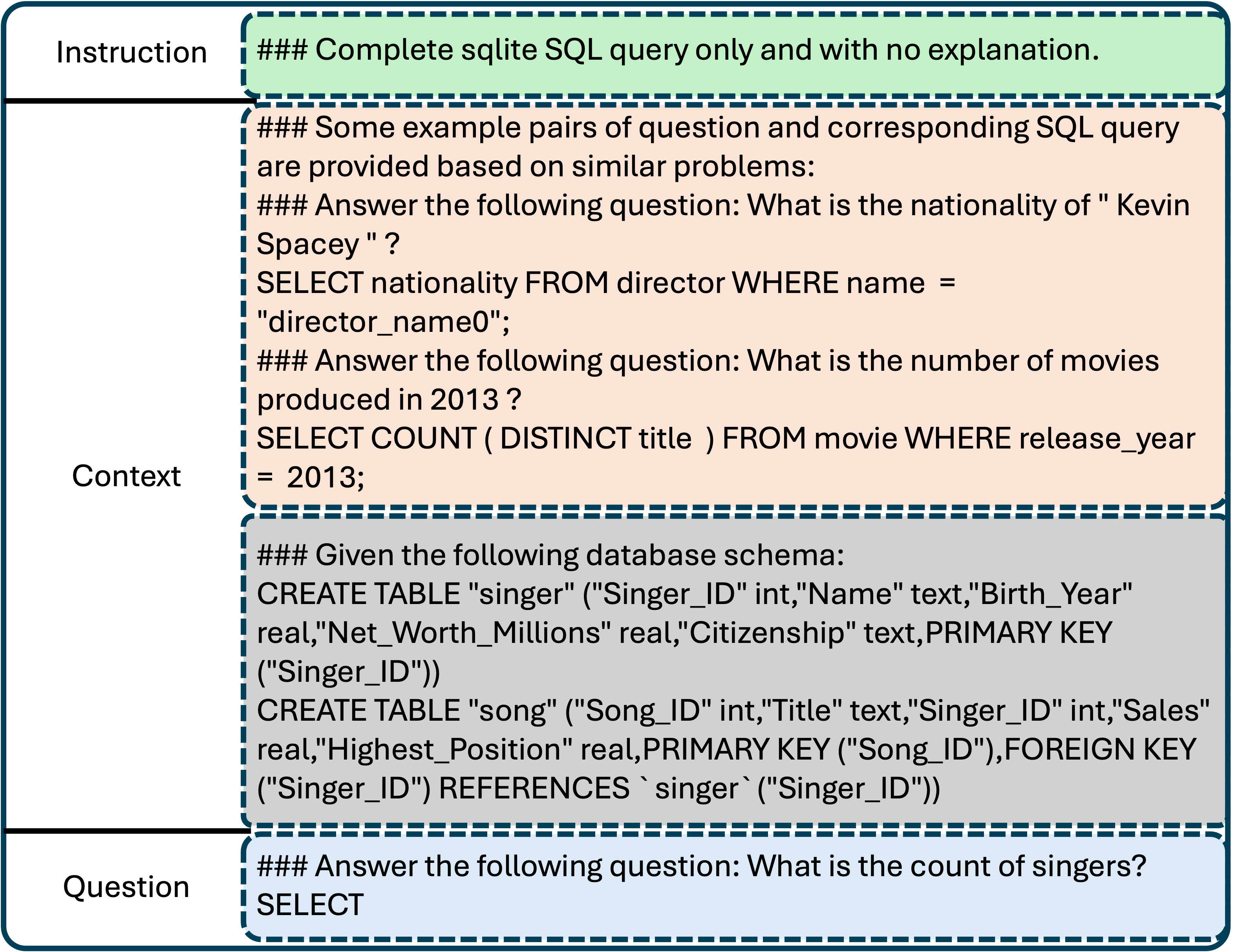}
	\caption{An Example of the Prompt Template}
	\label{fig:prompt}
\end{figure}

Given a natural language question $Q$ and a database $D$, the text-to-SQL problem aims to find an SQL query $Y$ that corresponds to the question.
The LLM-based solutions for text-to-SQL~\cite{DBLP:conf/emnlp/Nan0ZRTZCR23,DBLP:conf/nips/PourrezaR23,DBLP:journals/corr/abs-2307-07306,DBLP:journals/pvldb/GaoWLSQDZ24} formulate it as a generation problem that employs a prompt $P$ for the LLM $\mathcal{M}$.
It estimates the probability distribution over the potential SQL queries $Y$ and generates the SQL statement token-by-token.
This generation process could be formulated as Equation~\ref{eq-prb}:
\begin{equation} \label{eq-prb}
	{Pr}_{\mathcal{M}}(Y | P, D, Q) = \prod_{i = 1}^{|Y|} {Pr}_{\mathcal{M}}(Y_{i} | P, D, Q, Y[0...i-1])
\end{equation}
where $Y[0...i-1]$ is the sequence generated by the model so far before step $i$.

An example of a prompt template over LLM for the Text-to-SQL problem is shown in Figure~\ref{fig:prompt}, where the SQL generated by GPT-4 is:  \texttt{SELECT count(*) FROM singer}. 
The prompt for text-to-SQL typically includes three key components:
\begin{compactitem}
	\item \textbf{Instruction}, giving the general task descriptions.
	\item \textbf{Context}, providing the necessary context for task and demonstration examples.
	This is the most important component of a prompt.
	\item \textbf{Question}, describing the expected answer from (e.g. natural language question) 
\end{compactitem}

In this example, the prompt's context includes two parts: (1) \emph{Demonstration}, which provides some examples for few-shot learning, and (2) \emph{Schema}, which displays the schema information of the targeted database to offer hints for generating the SQL. 
As such, an essential goal is to provide high-quality context with an appropriate strategy to construct the prompt.

\section{Methodology}\label{sec-method}

\subsection{Overview}\label{subsec-overall}

\begin{figure}[h]
	\centering
	\includegraphics[width=0.45\textwidth]{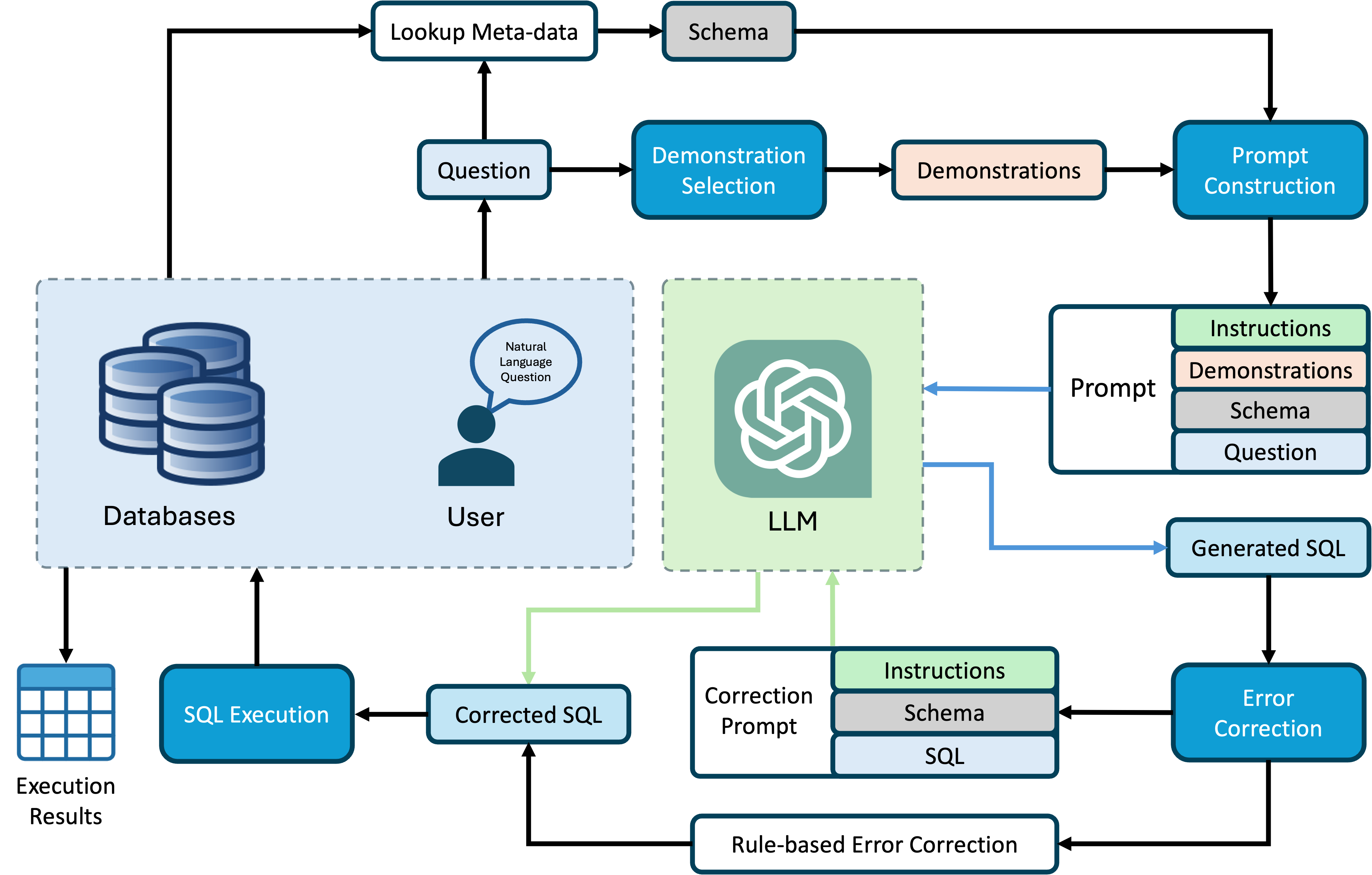}
	\caption{Overall Framework.}
	\label{fig:system}
\end{figure}

The overall framework of \name is shown in Figure~\ref{fig:system}. 
Given a question in natural language, we use the question to fetch (1) database schema and (2) demonstration examples to construct the prompt. 
The database schema suggests the necessary background specifics for SQL statements, such as the names and types of tables and attributes.  
Following the previous work ~\cite{DBLP:journals/pvldb/GaoWLSQDZ24}, we use the corresponding Data Definition Language (DDL) to describe the schema.
The demonstration examples provide useful contextual information to the LLM to facilitate the generation of SQL.
Following the discussions in previous works, we explored several options and proposed two new structure-based methods in Section~\ref{subsec-demo} and Section~\ref{subsec-gcl}, respectively. 
After obtaining the initial output SQL from LLM, we further perform error correction in the post-processing step to find and fix potential errors in Section~\ref{subsec-error}.

\subsection{Demonstration Selection Strategies: Basics}\label{subsec-demo}

In the process of prompt engineering over LLM, one essential way is through in-context learning~\cite{DBLP:conf/nips/BrownMRSKDNSSAA20} where LLM could use the contextual examples and condition its generation by recognizing patterns in the input.
This allows LLMs to perform new tasks during inference without any task-specific fine-tuning.
Previous studies~\cite{DBLP:conf/acl-deelio/LiuSZDCC22,DBLP:conf/acl/LuBM0S22} have shown that it is essential to select helpful examples to support the in-context learning with few-shot examples over LLM in different kinds of applications.
Similarly, it is also essential for prompt construction in the Text-to-SQL problem.

To this end, we focused on developing effective techniques to select demonstrations in the prompt construction process for the Text-to-SQL problem.
According to previous studies~\cite{DBLP:journals/pvldb/GaoWLSQDZ24,DBLP:conf/emnlp/Nan0ZRTZCR23}, it is essential to select examples that are similar to the given question instance.
Following this route, we address this problem by first defining a similarity metric to evaluate the relevance between a given instance and candidate examples and then selecting the top-ranked ones.
The candidates of demonstration examples could be pairs of a natural language question and the corresponding SQL that do not appear in the set of questions (e.g. dev and test sets in a benchmarking dataset). 
We will start from 3 basic approaches to select $k$ demonstration examples motivated by the high level idea in previous studies~\cite{DBLP:journals/pvldb/GaoWLSQDZ24,DBLP:conf/emnlp/Nan0ZRTZCR23}:

\noindent \textbf{Random.}\hspace{.5em} In this approach, $k$ examples are selected via random sampling from the available candidates. It is considered the baseline method in several previous studies.
\smallskip

\noindent \textbf{Hardness}\hspace{.5em} In this approach, we use query difficulty as a measure.
Examples are randomly selected from the group of instances that has the same level of hardness with question instance.
The Spider dataset~\cite{DBLP:conf/emnlp/YuZYYWLMLYRZR18} provides a tag of difficulty level (easy, medium, hard, extra) for each instance, and thus we can directly use it for selection.
For other datasets without such information in the metadata, we can use some rule-based heuristics to decide the hardness following the practice of previous works~\cite{DBLP:conf/emnlp/YuZYYWLMLYRZR18,DBLP:conf/emnlp/ZhongYK20}.
\smallskip

\noindent \textbf{Question Similarity}\hspace{.5em} This approach uses the string similarity between questions as the measure.
Here, we choose Jaccard Similarity as the metric and select results with top-k highest scores as the results.
Unlike the other two methods, the results of this method are deterministic since it does not involve random selection.
\smallskip

Although existing works have explored various strategies for selecting demonstrations to be included as few-shot examples, they have certain deficiencies in the domain of SQL (will be illustrated in our empirical observations in Section~\ref{subsec-more}).
Since an LLM generates the SQL based on the input prompt, it is essential to provide some examples with SQL that is similar to the expected output.
We developed a new demonstration selection strategy based on \emph{structure similarity} of SQL statements to address this issue.
To describe the structure of an SQL, we consider its Abstract Syntax Tree (AST), which consists of the relational operators.
AST is a general data structure that is independent of the specific database systems.
It is also a relatively lightweight data structure and could be generated without an underlying database system.
Specifically, we generate the trees through the third-party parsing tool sqlglot~\footnote{https://github.com/tobymao/sqlglot}.
In this tree-based solution, we focus on the structure information and ignore the exact names of tables, attributes, and predicates. 
Only the type information of tree nodes is kept as the \emph{node label}.
% This also helps achieve a domain-agnostic solution to the Text-to-SQL problem that could be easily adopted to databases with new domains.
Examples of node label include SELECT, WHERE, TABLE, aggregation operations (e.g. MIN, MAX, GROUP BY), conjunctions (e.g. INTERSECTION, EXCEPT, UNION), HAVING, and ORDER BY etc. 

After transforming the SQL into the above tree structure, we then evaluate the structure similarity.
Here, we choose \emph{tree edit distance}~\cite{DBLP:journals/tcs/Bille05} as the similarity metric.
Basically, given two labeled trees, tree edit distance is the minimum number of edit operations that is needed to transform one tree into another.
There are three kinds of edit operations: 
\begin{compactitem}
	\item Insertion: insert a node between an existing node and a subsequence of consecutive children of this node;
	\item Deletion: delete a node and connect its children to its parent, maintaining the order;
	\item Substitution: rename the label of a node.
\end{compactitem}
We will select examples with the top-k smallest tree edit distance from the question instance as the demonstration.

\begin{figure}[ht]
	\centering
	\begin{subfigure}{\columnwidth}
		\centering
		\includegraphics[width=0.85\textwidth]{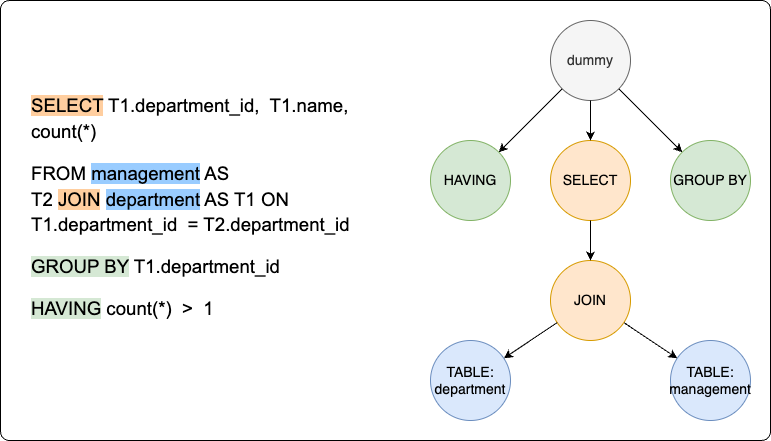}
	\end{subfigure}
	\begin{subfigure}{\columnwidth}
		\centering
		\includegraphics[width=0.85\textwidth]{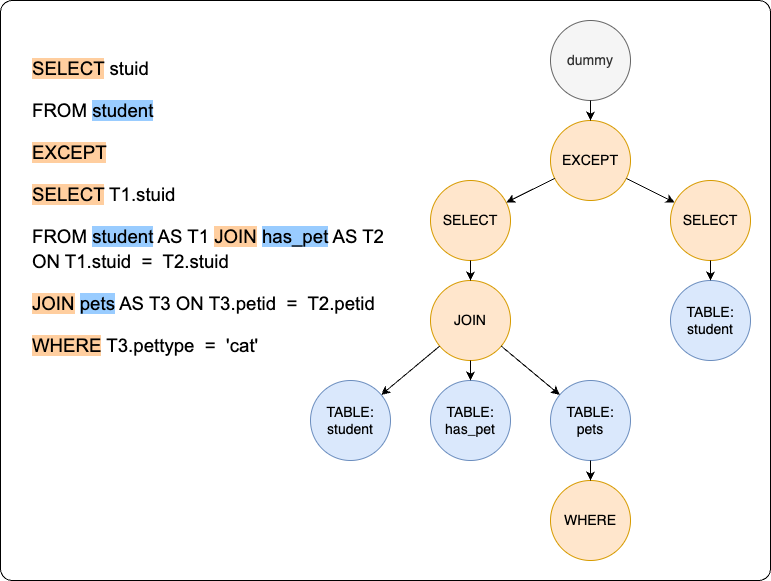}
	\end{subfigure}
	\caption{Examples of Tree Edit Distance between ASTs of SQL. Names of tables are kept just for reference but are not considered as node labels in measurement}
	\label{fig:treesim}
\end{figure}

\begin{example}
	An example of using tree edit distance to evaluate the structure similarity between SQL clauses is shown in Figure~\ref{fig:treesim}.
	We transformed the two SQL clauses into the AST and used different colors to denote different node labels.
	To transform the AST of the first SQL into that of the second one, we need the following operations: 3 deletions on the HAVING, GROUP BY, and TABLE: department nodes, 1 substitution to replace the SELECT with EXCEPT, 6 insertions to insert the left subtree of the EXCEPT node.
	As such, the tree edit distance is 10 in total. 
	In this example, these two instances are not similar and should not be considered as a demonstration example for each other.
\end{example}

Although tree edit distance could accurately reflect the structure similarity between SQL clauses, its computational cost is $\mathcal{O}(n^3)$, where $n$ is the number of nodes in the tree.
Since the candidate set of examples could be very large, computing the tree edit distance between the given instance and all potential candidate examples is very expensive.
To address this problem, we adopt the idea of pq-gram from a previous study~\cite{DBLP:journals/tods/AugstenBG10} to compute an estimation instead of the exact value of the tree edit distance.
It is a signature that can help estimate the tree edit distance between two trees with less cost.
To this end, we must obtain a set of pq-grams for each tree.
Suppose the set of pg-grams for two trees is $L_1$ and $L_2$, respectively; the \emph{pq-gram distance} between the two trees can be calculated as $|L_1 \cup L_2| - 2*|L_1 \cap L_2|$.
The pq-gram distance could be calculated in $\mathcal{O}(n \log n)$ time and serves as a lower bound of tree edit distance.
And we will use the pq-gram distance between two ASTs instead of actual tree edit distance to select the demonstration examples.
Due to the space limitation, here we omit the details of computation and proof of correctness, which could be found in~\cite{DBLP:journals/tods/AugstenBG10}.

One remaining issue to be resolved is that when constructing the prompt for a question, we only have the natural language question but not the actual SQL.
Our solution is to first conduct a prompt with zero-shot learning to generate an initial SQL and use it as the query to find structurally similar examples.
The extra overhead would be trivial since the prompt for zero-shot learning is much shorter than that with demonstration examples. 

\subsection{Graph based Demonstration Selection}\label{subsec-gcl}

While the above tree-based solution could capture the structural information of SQL statements, it still did not consider some important information, such as predicate values and column names in the involved tables.
In addition, the structural similarity is evaluated by tree edit distance, which is based on syntactic similarity and thus might lose some latent structural information.
In this section, we propose a graph-based demonstration selection approach to address such issues and further improve performance.
Compared with the tree structure, the graph can carry not only richer structural information but also additional semantics.
The basic idea is to construct a directed acyclic graph (DAG) to represent each SQL statement. 
In this way, the similarity between two SQL statements could be evaluated by that between their corresponding DAGs.
% Then given a query instance, we will select examples with top-$k$ similar DAGs.

\begin{figure}[ht]
	\centering
	\includegraphics[width=0.5\textwidth]{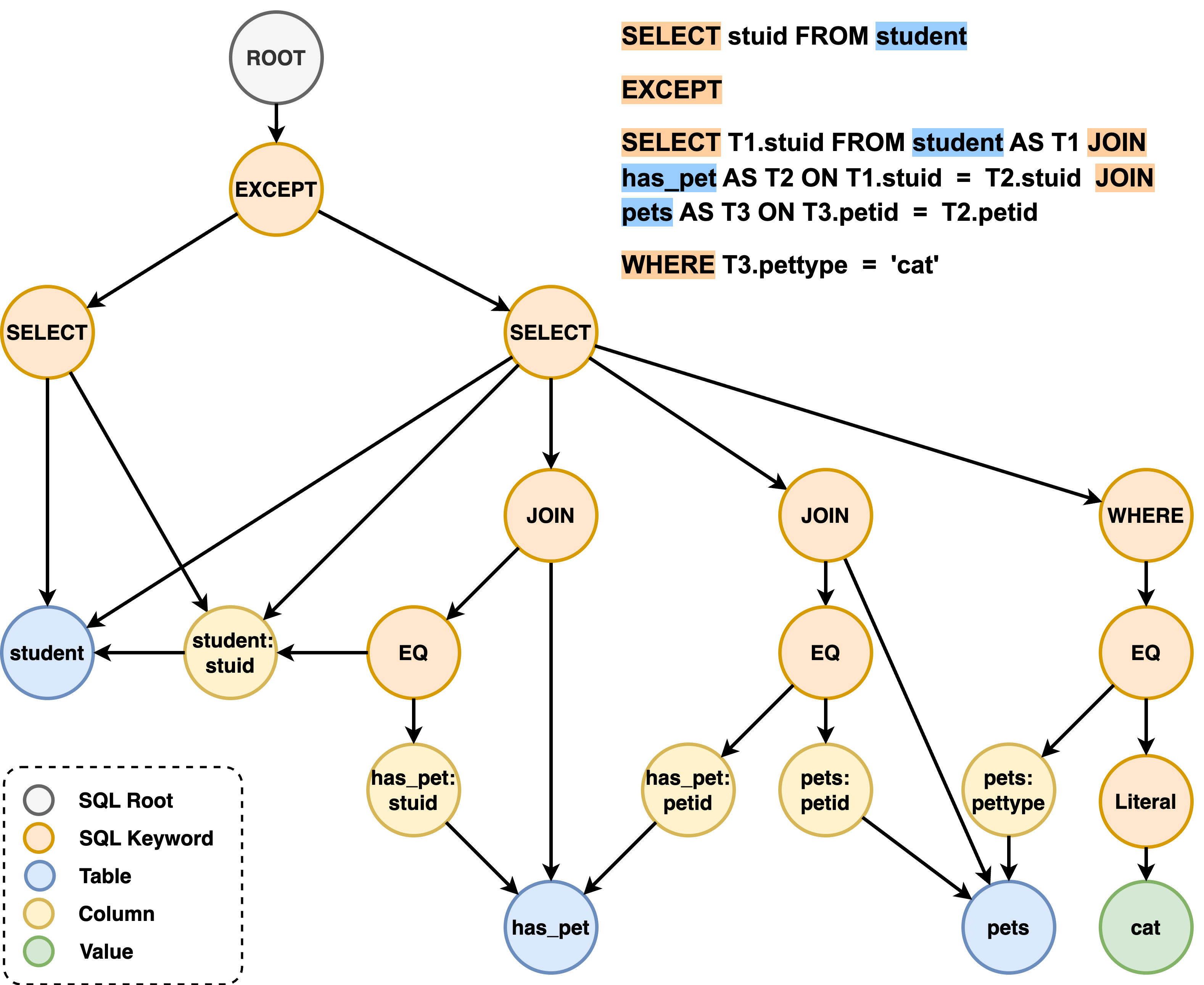}
	\caption{An Example of a SQL Statement and its Graph Representation.}
	\label{fig:graph}
\end{figure}

To reach this goal, the first step is to construct the graph (DAG) from a SQL statement.
We extend the Abstract Syntax Tree (AST) representation described above by incorporating additional information. 
The graph consists of five types of node labels: (dummy) Root, SQL Keyword, Table, Column, and Value. 
Each unique SQL keyword is represented as an individual node (e.g., multiple \texttt{JOIN} nodes for a query). In contrast, identical table or column names are merged into a single node to maintain subgraph connectivity.
Beyond the basic AST structure illustrated in Figure~\ref{fig:treesim}, we also include table and column names, as well as predicate values in the SQL query. 
After defining the nodes, we then add edges to capture relationships between them based on the following rules: (1) between each operator and the associated table or column name, (2) between each column and the table it belongs to, and (3) between a predicate value and its corresponding literal operator.
Additionally, SQL keywords like \texttt{HAVING} and \texttt{GROUP BY} are connected to their parent \texttt{SELECT} nodes. 
An example of representing SQL with the DAG is shown in Figure~\ref{fig:graph}.
Here, we first obtain the tree structure, which consists of essential operators in the SQL query.
Next, we further parse the predicates and identify the columns (yellow nodes), literals (e.g. EQ), and values (green nodes).
Finally, we add edges between such newly created nodes and the existing nodes corresponding to operators and obtain the graph.

Given such graph structures, we could evaluate the similarity based on cosine similarity between graph embeddings.
To this end, we need to train a node encoder for the SQL graph.
However, there is no labeled training instance, and the training process needs to be conducted in a fully unsupervised manner.
To satisfy such needs, we employ the technique of graph contrastive learning~\cite{DBLP:conf/iclr/SunHV020,DBLP:conf/nips/YouCSCWS20} as the solution.
Graph contrastive learning is a variant of self-supervised learning that enables the training of graph encoders, such as Graph Neural Networks (GNNs), without human annotations.
This is realized by constructing multiple graph views via stochastic augmentation of the input graph and then learning representations by contrasting positive samples against negative ones~\cite{DBLP:conf/nips/0001XLW21}.
As illustrated in previous studies, two important factors of graph contrastive learning are contrastive instances and contrastive objective.
While we can continue to employ the loss function from the previous studies as the objective, we need to consider the semantics of SQL when defining the graph augmentation operations for creating contrastive instances.
According to our definition of the SQL graph, if we directly apply operations from previous studies~\cite{DBLP:conf/iclr/SunHV020,DBLP:conf/nips/YouCSCWS20}, we might end up with a result that is corresponding to a totally different SQL statement or even an invalid one.
For example, if the node corresponding to a JOIN operation is replaced with a SELECT one, although the topological structure of a SQL might still be close to the original instance, the structure of the corresponding SQL query will change greatly.

To keep the basic semantics of SQL statements when creating the contrastive instances, we define the following operations to perform augmentation of the input graph.\smallskip

\noindent \textbf{Feature Masking}\hspace{.5em} This operator randomly masks the node feature with a \texttt{$<$MASK$>$} token in LLM; while nodes with essential keywords (\texttt{ROOT}, \texttt{SELECT}, \texttt{JOIN}, \texttt{WHERE}, \texttt{GROUP}, \texttt{ORDER}) will not be masked.\smallskip

\noindent \textbf{Keyword Replacement}\hspace{.5em} This operator selects the SQL keywords that can be replaced while still keeping a valid SQL. These SQL keywords include logical and comparison operators (e.g., \texttt{EQ}, \texttt{AND}), arithmetic operators (e.g., \texttt{ADD}, \texttt{DIV}), and aggregations (e.g., \texttt{COUNT}, \texttt{SUM}, \texttt{MIN}). Each selected keyword node will be randomly replaced with a valid keyword node of the same type. For example, \texttt{GT} ($>$) could be replaced by \texttt{LT} ($<$), \texttt{GTE} ($\geq$) and \texttt{LTE} ($\leq$).\smallskip

\noindent \textbf{Value Replacement}\hspace{.5em} This operator selects nodes with type "VALUE", then replace it with new random values having the same data type (BOOLEAN, INT, FLOAT, STRING). Especially, for values that refer to partial match in SQL (e.g., "\%USA"), only the partial string will be replaced (e.g., "\%USA" to "\%Canada").\smallskip

\noindent \textbf{Database Replacement}\hspace{.5em} This operator replaces the whole tables and columns that belong to one database with tables and columns in another database. It prefers to select new columns that have the same column type (e.g., numerical) as the original one to ensure that the augmented graph is a valid SQL.\smallskip

\noindent \textbf{Predicate Modification}\hspace{.5em} This operator chooses the predicate (e.g., \texttt{WHERE}, \texttt{HAVING} clause) of a SQL statement and then either randomly drops either the entire predicate or simplifies the condition in the predicate (e.g. "\texttt{WHERE A=1 AND B=2}" to "\texttt{WHERE B=2}"). \smallskip

\noindent \textbf{Join Simplification}\hspace{.5em} 
For \texttt{SELECT} nodes with more than one \texttt{JOIN} node as neighbors, this operator randomly drops one \texttt{JOIN} node and corresponding clause. If there are nodes (e.g., TABLE or COLUMN) in such clause that are also connected to other nodes associated with essential SQL keywords, they will be kept.\smallskip

\begin{figure}[ht]
	\centering
	\includegraphics[width=0.5\textwidth]{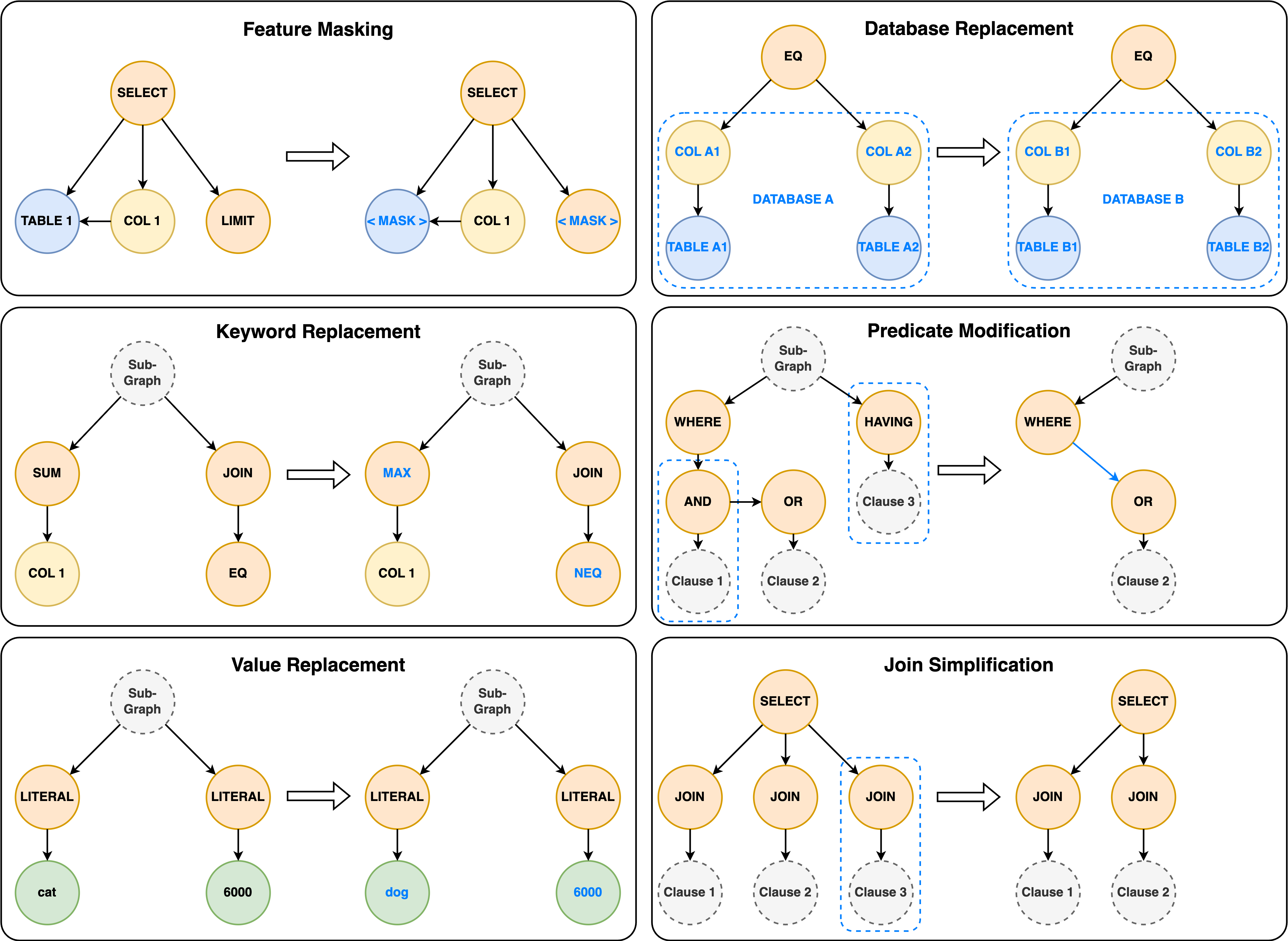}
	\caption{Examples for Graph Augmentation Operators. The blue font highlights the operations}
	\label{fig:graphop}
\end{figure}

\begin{example}
	We provide examples of the above-defined graph operations in Figure~\ref{fig:graphop}.
	For the SQL graph in Figure~\ref{fig:graph}, the Feature Masking operator might randomly mask column node  \texttt{student:stuid} and \texttt{EQ} keyword to \texttt{$<$MASK$>$} tokens. 
	Keyword Replacement operator might replace the \texttt{EQ} under second \texttt{JOIN} with \texttt{NEQ}. 
	Value Replacement operator might replace value node \texttt{cat} with \texttt{bird}.
	Database Replacement operator might replace table nodes \texttt{student}, \texttt{has\_pet}, \texttt{pets} with \texttt{singer}, \texttt{singer\_in\_concert}, \texttt{concert}, respectively, and replace column nodes \texttt{student:stuid}, \texttt{has\_pet:stuid}, \texttt{has\_pet:petid}, \texttt{pets:petid} and \texttt{pets:pettype} with \texttt{singer:Singer\_ID}, \texttt{singer\_in\_concert:Singer\_ID}, \texttt{singer\_in\_concert:concert\_ID}, \texttt{concert:concert\_ID} and \texttt{concert:Theme}, respectively. 
    Predicate Modification operator might drop \texttt{WHERE} node and its successors (\texttt{pets} will not be dropped because other parts also use it in the graph). 
    The JOIN Simplification operator might drop the second \texttt{JOIN} node and its successors.
\end{example}

With the help of such operators, we are then able to obtain the contrastive instances.
Given an original instance, we will randomly apply an operation defined above over it to obtain a positive instance;
Meanwhile, the negative instances could be obtained by randomly sampling from the rest of the instances.

\begin{example}
We provide an example of the above approaches of generating positive and negative instances for graph contrastive learning. 
Given the DAG for SQL shown in Fig~\ref{fig:graph}, one positive instance could be obtained by applying the Predicate Modification operator to drop the where clause. 
Consequently, the corresponding SQL statement is "\texttt{SELECT stuid FROM student EXCEPT SELECT T1.stuid FROM student AS T1 JOIN has\_pet AS T2 ON T1.stuid = T2.stuid JOIN pets AS T3 ON T3.petid = T2.petid}", which shares a similar structure with the original one. 
Meanwhile, a negative instance could be randomly selected from the rest of the dataset, an example could be the DAG corresponding to the SQL "\texttt{SELECT Theme FROM farm\_competition ORDER BY YEAR ASC}".
\end{example}

Next, we employ the instances created with above approaches to train a graph encoder.
In this part, the graph encoder can be implemented using various Graph Neural Network (GNN) architectures.
In our implementation, we utilize a 2-layer Graph Attention Network (GAT) as the encoder. 
We denote a graph as $G = \langle V, E, L \rangle$, where $V = \{ v_i \}_{i=1}^n$ and $E \subseteq V \times V$ denotes the set of nodes and edges, respectively.
And $L$ denotes the labeling function that assigns a label to each node.
Each node $v$ is initialized with concatenated features combining one-hot encoding of its node label $\mathbf{l}_v$. and text embedding $\mathbf{e}_v$ as illustrated in Equation~\ref{eq-node}.
\begin{equation}\label{eq-node}
\mathbf{h}_v^{(0)} = \text{CONCAT}(\mathbf{l}_v, \mathbf{e}_v)
\end{equation}
Here, we obtained the text embedding by encoding the texts associated with a node, e.g., SQL keyword, column name, value, etc, with the pre-trained SentenceBert~\cite{DBLP:conf/emnlp/ReimersG19} model.
With this node representation, we compute the propagation of representation at GNN layer $k$ as Equation~\ref{eq-lk}:
\begin{equation}\label{eq-lk}
\mathbf{h}_v^{(k+1)} = \text{AGG}\left(\{\text{COMBINE}(\mathbf{h}_v^{(k)}, \mathbf{h}_u^{(k)}) \mid u \in \mathcal{N}(v)\}\right))
\end{equation}
where $\mathbf{h}_v^{(k)}$ is the node representation at layer $k$; $\text{AGG}_{(k)}(\cdot)$ is the aggregation function that aggregates the information from neighbor nodes during the message-passing process; 
$\text{COMBINE}_{(k)}(\cdot)$ is the function to merge node features with features aggregated from neighbors in the GNN layer;
and $\mathcal{N}(v)$ denotes the set of neighbors of node $v$ in the graph. 

After obtaining each node embedding in the above method,  we use graph readout function~\cite{DBLP:conf/iclr/SunHV020} to generate the graph embedding $\mathbf{h}_G$ by aggregating that of all nodes as shown in Equation~\ref{eq-readout}:
\begin{equation}\label{eq-readout}
\mathbf{h}_G = \text{READOUT}\left(\{\mathbf{h}_v^{(n)} \mid v \in \mathcal{V}\}\right)
\end{equation}
where the READOUT layer combines mean, sum, and max aggregations over node embeddings in our implementation.

Then, the final graph embedding is obtained by adding a two-layer MLP projection head on top of the aggregated node embeddings as illustrated in Equation~\ref{eq-output}:
\begin{equation}\label{eq-output}
\mathbf{z}_G = \text{MLP}(\mathbf{h}_G)
\end{equation}

The contrastive loss maximizes the similarity between an anchor graph and its positive samples while minimizing similarity with negative samples. We used normalized temperature-scaled cross-entropy loss NT-Xent widely utilized in previous studies~\cite{DBLP:conf/nips/Sohn16,DBLP:journals/corr/abs-1807-03748}.
The details  are shown in Equation~\ref{eq-loss}:

\begin{equation}\label{eq-loss}
\mathcal{L}_i = - \log \frac{\sum_{j=1}^{n_{\text{positive}}} \exp\left( \frac{\text{sim}(\mathbf{z}_i, \mathbf{z}_j^+)}{\tau} \right)}{\sum_{j=1}^{n_{\text{positive}}} \exp\left( \frac{\text{sim}(\mathbf{z}_i, \mathbf{z}_j^+)}{\tau} \right) + \sum_{k=1}^{n_{\text{negative}}} \exp\left( \frac{\text{sim}(\mathbf{z}_i, \mathbf{z}_k^-)}{\tau} \right)}
\end{equation}
where $\mathbf{z}_i$, $\mathbf{z}_j^+$ and $\mathbf{z}_k^-$ is the embedding of the anchor graph, the positive graph for the anchor and negative graph for the anchor, respectively; 
$n_{\text{positive}}$ and $n_{\text{negative}}$ is the number of positive graphs and negative graphs per anchor, respectively; 
$\text{sim}(\cdot)$ is the cosine similarity between between two embeddings; 
$\tau$ is the temperature parameter for scaling the similarity scores.

\subsection{Error Correction}\label{subsec-error}

Although LLMs are powerful in generating SQL statements based on input questions, some output statements might still be invalid due to a large training corpus beyond SQL and data that is not strictly compliant with SQL syntax.
Potential errors also exist due to a lack of understanding of the contextual information or the question in the prompt.
To address such issues, we propose an error correction module that automatically fixes such errors in the post-processing step. 
We proposed two kinds of error correction methods: rule-based and prompt-based.

First of all, we develop a set of rules to fix some simple errors based on the efforts of analyzing typical mistakes, an incomplete list of examples is as follows
% \eser{examples could be useful}
% Jin: here we just listed some examples of rules but not all of them. If you mention

\noindent\textbf{String Format.}\hspace{.5em} Sometimes, the structure of generated SQL aligns with the ground truth, but there are mismatches between the values in the predicates, resulting in different execution results with the golden SQL.
If such a mismatch is caused by string format issues such as spelling and cases, we can fix it via rules that align the values in the generated SQL with those in the database.
\smallskip

\noindent\textbf{Mismatch in Schema.}\hspace{.5em} The LLM might involve non-existent or incorrect names of tables and attributes in the output due to hallucination.
We will look up the metadata to ensure all the table and attribute names exist.
If we find non-existing ones, we replace them with the most similar ones from the metadata to ensure the generated SQL is valid.
\smallskip

\noindent\textbf{Invalid Aggregation.}\hspace{.5em} We will fix the invalid aggregations, such as MIN and MAX, over non-numerical attributes or COUNT on multiple attributes.
For the former case, we will directly remove it from the generated SQL; 
For the latter case, we will replace the attribute with the first attribute or *.
\smallskip

\noindent\textbf{Join Condition.}\hspace{.5em} If the join condition happens between keys that are not joinable, we will replace it with foreign keys that are joinable between the two tables. 
If that does not exist, we will remove the join condition.
\smallskip 

In the above process, we look into the database to fix the errors related to string format and minor syntax issues such as upper/lower cases but do not consider the semantics of contents.
Therefore, we do not use database contents to facilitate the semantic understanding of the question or SQL generation.
Some previous works~\cite{DBLP:journals/pvldb/GaoWLSQDZ24,DBLP:journals/corr/abs-2307-07306} also employ self-consistency techniques~\cite{DBLP:conf/iclr/0002WSLCNCZ23} for post-processing, which needs to execute the generated SQL in database before making the final output.
Unlike such practices, we did not utilize the execution results of SQL in our approach.

\begin{figure}[ht]
	\centering
	\includegraphics[width=0.45\textwidth]{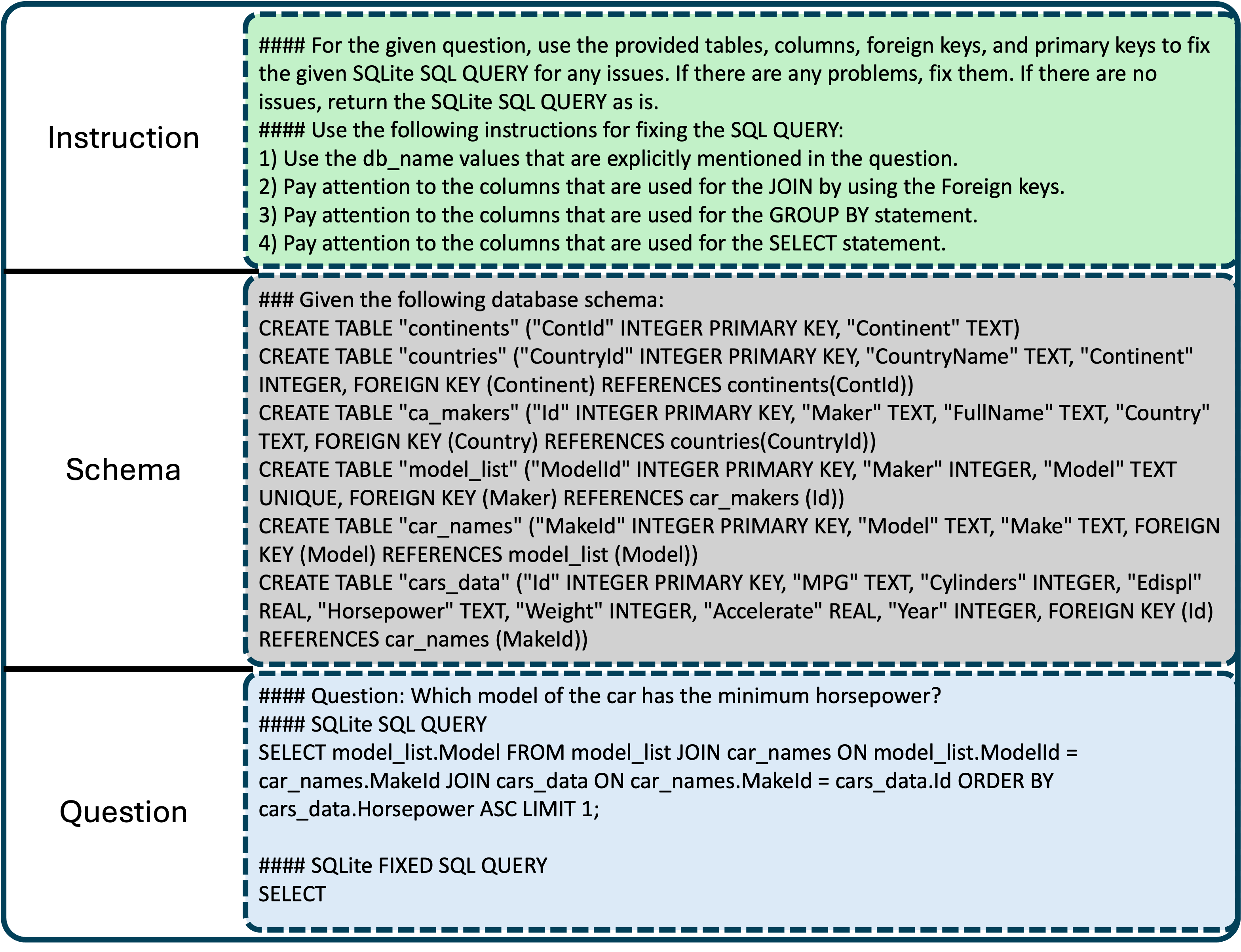}
	\caption{The Template for Prompt-based Error Correction}
	\label{fig:error}
\end{figure}

We also develop a prompt-based method to correct errors in the generated SQL with one more iteration with LLMs in a zero-shot learning manner.
An example is shown in Figure~\ref{fig:error}.
The structure of this prompt is similar to that shown in Figure~\ref{fig:prompt}, i.e., it will include the instruction and schema information.
In addition, the generated SQL is also included as part of the question, and the request is to ask LLM to correct the potential errors.
In this process, we provide some guidelines in the format of explicit rules as hints for the LLM to make proper corrections, such as ``Pay attention to whether every join condition is necessary'' and ``Use DESC and DISTINCT when needed''.
The choice of guidelines could be realized by rule-based heuristics. 
For example, if the required contents could be selected from an original table without a join, we will apply the guideline related to join conditions.
In the above example shown in Figure~\ref{fig:error}, we assume the second rule related to the join condition could help the LLM to recognize the unnecessary join condition in the generated SQL and fix it via another prompt.

It is easy to see that the rule-based method is simple but has limited coverage.
At the same time, prompt-based methods are expensive as they require another generation, but they could fix some complicated errors with well-designed instructions.
To make good use of both, we develop a choice strategy to decide whether to use each of them in the following way:
First, we apply all rule-based methods for correction. 
If they can find some errors and fix them, we will not continue applying prompt-based methods.
In addition, since the prompt-based method aims to fix complicated errors, we only apply it when the case seems to be challenging.
For example, in the Spider dataset~\cite{DBLP:conf/emnlp/YuZYYWLMLYRZR18}, we apply the prompt-based method only for instances belonging to hard and extra categories.
When such information is missing, we can look into the demonstration examples in the original prompt, e.g., if there are examples with join conditions between more than 2 tables or conjunction operations, we will apply the prompt-based method.

\section{Evaluation}\label{sec-exp}

\subsection{Experiment Setup}\label{subsec-setup}
\vspace{-1em}
\begin{table}[ht]
	\centering
	\caption{The statistics of datasets}\label{tbl-stats}
	\begin{tabular}{l|ccc}
		\toprule
		Dataset & \# Queries & \# Databases & \# Tables \\
		\midrule
		Spider (train) & 8,659 & 146 & 795 \\
		Spider (dev) & 1,034 & 20 & 81  \\
		Spider (test) & 2,147 & 40 & 180 \\
		BIRD (train) & 9,428 & 69 & 524 \\
		BIRD (dev) & 1,533 & 11 & 81 \\
		\bottomrule
	\end{tabular}
\end{table}

\subsubsection{Datasets}

We mainly conducted experiments on the benchmarking dataset Spider~\cite{DBLP:conf/emnlp/YuZYYWLMLYRZR18}, which is widely used in previous studies about Text-to-SQL.
We reported results on dev and test sets, which are released on the official website~\footnote{https://yale-lily.github.io/spider}.
The instances in the training set of Spider are used as the candidate of demonstration examples for few-shot learning and report the results on both the dev and test sets.
In addition, we also evaluated on the BIRD~\cite{DBLP:conf/nips/LiHQYLLWQGHZ0LC23} dataset, which is recently proposed to evaluate the efficiency of the generated SQLs. 
Since our work focused on improving the effectiveness rather than efficiency of Text-to-SQL tasks, we only report results regarding accuracy but not the Valid Efficiency Score, which evaluates whether the generated SQL queries are optimized.
The detailed statistics of these datasets are shown in Table~\ref{tbl-stats}.

\subsubsection{Evaluation Metrics}

Following the practice of previous studies~\cite{DBLP:conf/emnlp/ZhongYK20}, we use two metrics to evaluate the accuracy of the proposed solutions:
Exact Set Match (EM) and Execution Match (EX) accuracy.
Exact Set Match accuracy, which is also known as logical form accuracy, measures the matched SQL keywords between the predicted SQL query and the corresponding ground truth.
Execution Match accuracy requires executing the generated SQL in a real database system and comparing the execution result with that of the gold standard SQL. 
It provides a more precise estimate of the model’s performance since multiple valid SQL statements may exist for a single question.
Furthermore, we also report the cost on Spider datasets by examining the total number of tokens in the prompts.

\subsubsection{Baseline Methods}

We primarily choose the following existing solutions as baseline methods to compare with:

\noindent \textbf{DAIL-SQL}~\cite{DBLP:journals/pvldb/GaoWLSQDZ24} is the latest prompt-based method that explores a wise combination of prompt template and demonstration selection methods to improve the overall performance of LLMs.
\smallskip

\noindent \textbf{DIN-SQL}~\cite{DBLP:conf/nips/PourrezaR23} utilizes a chain-of-thought strategy to divide the text-to-SQL problem into 3 stages and conduct prompts for each of them, respectively.
\smallskip

\noindent \textbf{Augment}~\cite{DBLP:conf/emnlp/Nan0ZRTZCR23} proposes a schema-related knowledge augmentation method to improve the prompt construction process to obtain high-quality SQL based on LLM.
\smallskip

\noindent \textbf{CatSQL}~\cite{DBLP:journals/pvldb/FuLWLTS23} is a template-filling based method that achieves the best performance in that category of works.
\smallskip

\noindent \textbf{Graphix-T5}~\cite{DBLP:conf/aaai/LiHCQ0HHDSL23} constructs a graph to model the interaction between the question and database schema and then incorporates such information in the fine-tuning process of the decoder. 
It is the up-to-date one in the category of machine-translation-based methods.
\smallskip

Besides, we also include the results of earlier representative works, such as PICARD~\cite{DBLP:conf/emnlp/ScholakSB21}, RASAT~\cite{DBLP:conf/emnlp/QiTHW0ZWZL22}, RYANSQL~\cite{DBLP:journals/coling/ChoiSKS21}, LGESQL~\cite{DBLP:conf/acl/CaoC0ZZ020}, SmBoP~\cite{DBLP:conf/naacl/RubinB21} and RESDSQL~\cite{DBLP:conf/aaai/LiHCQ0HHDSL23} in the comparison.
We directly cite the numbers from the original papers and the leader board for all methods.

\subsubsection{Environment}

We implemented all proposed methods in Python.
All experiments are run on a server with configurations similar to those of a g5.12xlarge AWS EC2 machine, which has one AMD EPYC 7R32 48-core processor and 192GB RAM. 
We reported the results of prompt over OpenAI APIs for both GPT-4 (gpt-4-0613) and GPT-3.5 (gpt-3.5-turbo-0125) for the SQL generation. 
Due to the budget constraint, we only use GPT-3.5 for the experiments with large-scale prompts, e.g., the study about different numbers of demonstrations in Figure~\ref{fig:dnum}.
Otherwise, the results for all LLM-based solutions will be based on GPT-4 if there is no additional explanation.

%The model was trained on the training split for 30 epochs, with a learning rate of 0.001 and a linear weight decay of $1 \times 10^{-4}$.

\subsection{Comparative Performance Measurement}\label{subsec-main}

\begin{table}[ht]
	\centering
	\caption{Main Results on the Spider Dataset. ``-'' means the corresponding result is not available in the original paper or any public leader board.}\label{tbl-main}
	\begin{tabular}{l|cc|cc}
	\toprule
	Method & \multicolumn{2}{c|}{Dev} & \multicolumn{2}{c}{Test} \\
	& EM (\%) & EX (\%)  & EM (\%) & EX (\%)  \\
	\midrule
	RYANSQL & 66.4 & 58.2 & - & - \\
	LGESQL & 75.1 & 34.8 & 72.0 & - \\
	SmBoP & 74.7 & 77.9 & 71.1 & 69.5 \\
	PICARD & 75.5 & 79.3 & - & 75.1 \\
	RASAT & 74.7 & 80.5 & 70.6 & 75.5 \\
	Graphix-T5 & 77.1 & 81.0 & \textbf{74.0} & 77.6 \\
	RESDSQL & 80.5 & 84.1 & 72.0 & 79.9 \\
	CatSQL & \textbf{80.6} & 83.7 & 73.9 & 78.0 \\
	\midrule 
	Augment  & - & 84.1 & - & - \\
	DIN-SQL  & 60.1 & 74.2 & 60.0 & 85.3 \\
	DAIL-SQL & 71.9 & 82.4 & - & 86.2 \\
	\midrule
	\name (GPT-3.5) & 58.5 & 81.6 & 55.7 & 80.5 \\
	\name & 69.7 & \textbf{87.4} & 67.2 & \textbf{86.8} \\
	\bottomrule
	\end{tabular}
\end{table}

The main results on the Spider dataset are shown in Table~\ref{tbl-main}.
We have the following observations: 
First of all, LLM-based solutions achieved better performance in EX. 
This is due to the power of LLM in understanding the input question and generating SQL accordingly.
At the same time, the results of EM are not as good as PLM-based methods.
The reason is that LLM-based solutions generate the SQL according to the semantics of the question based on the inherited knowledge gained in the pre-training process, while PLM-based methods learn the syntax of SQL queries from the training set, which has a syntax structure more similar to those in the dev and test sets.
Nevertheless, as shown in recent studies~\cite{DBLP:conf/aaai/Li00023,DBLP:conf/aaai/LiHCQ0HHDSL23,DBLP:conf/emnlp/Nan0ZRTZCR23,DBLP:journals/pvldb/FuLWLTS23,DBLP:conf/nips/PourrezaR23,DBLP:journals/pvldb/GaoWLSQDZ24}, EX is a more critical metric in evaluating the main results as it is more closely related to the performance in real scenarios.
Therefore, it is safe to claim the superiority of LLM-based methods only based on the EX results.

In addition, \name performs better than other LLM-based solutions~\footnote{DAIL-SQL with self-consistency could reach the results of 86.6. However, it requires running the prompt multiple times and executing the generated SQL in the database before the final output. We report the result without self-consistency for a fair comparison.}.
The reason is that \name proposed effective demonstration selection techniques that could customize the demonstration examples for each question.
In this way, it would provide useful signals for different questions to the LLM.
At the same time, our error correction techniques could help fix various errors in the LLM output.
In this way, errors due to lack of sufficient context information could be avoided.

\begin{table}[ht]
	\caption{Performance Breakdown based on Difficulty on Spider (EX \%)}\label{tbl-break}
	\scalebox{0.9}{
	\begin{tabular}{lc|ccccc}
		\toprule
		 Split & Method & Easy & Normal & Hard & Extra & Overall \\
		 \midrule
		 \multirow{2}{*}{Dev} & \name (GPT-3.5) & 93.1 & 86.8 & 70.7 & 62.0 & 81.6 \\
		 & \name & \textbf{96.4} & \textbf{90.8} & \textbf{82.2} & \textbf{70.5} & \textbf{87.4} \\
		 & CatSQL & 95.6 & 88.3 & 74.7 & 62.7 & 83.7 \\
		 & DIN-SQL & 91.1 & 79.8 & 64.9 & 43.4 & 74.2 \\
		 \midrule
		 \multirow{2}{*}{Test} & \name (GPT-3.5) & 91.9 & 83.5 & 71.7 & 69.5 & 80.5 \\
		 & \name & 92.3 & 89.6 & 82.1 & 78.7 & 86.8 \\
		 \bottomrule
	\end{tabular}
} 
\end{table}

We show the results of performance breakdown based on the query difficulty in Table~\ref{tbl-break}.
Since very few previous studies reported such results, we only include the comparison with CatSQL~\cite{DBLP:journals/pvldb/FuLWLTS23} and DIN-SQL~\cite{DBLP:conf/nips/PourrezaR23} on the dev set.
We can see that compared with previous studies, \name has more improvement in the harder cases.
Compared with the template filling-based method CatSQL, \name could take advantage of the power of LLM in understanding the question and code generation to improve the overall performance.
The performance of \name is much better than another LLM-based method DIN-SQL in hard and extra-hard categories.
The reason might be the useful insights provided by properly selected demonstrations.

\begin{figure}[h]
	\centering
	\includegraphics[width=0.5\textwidth]{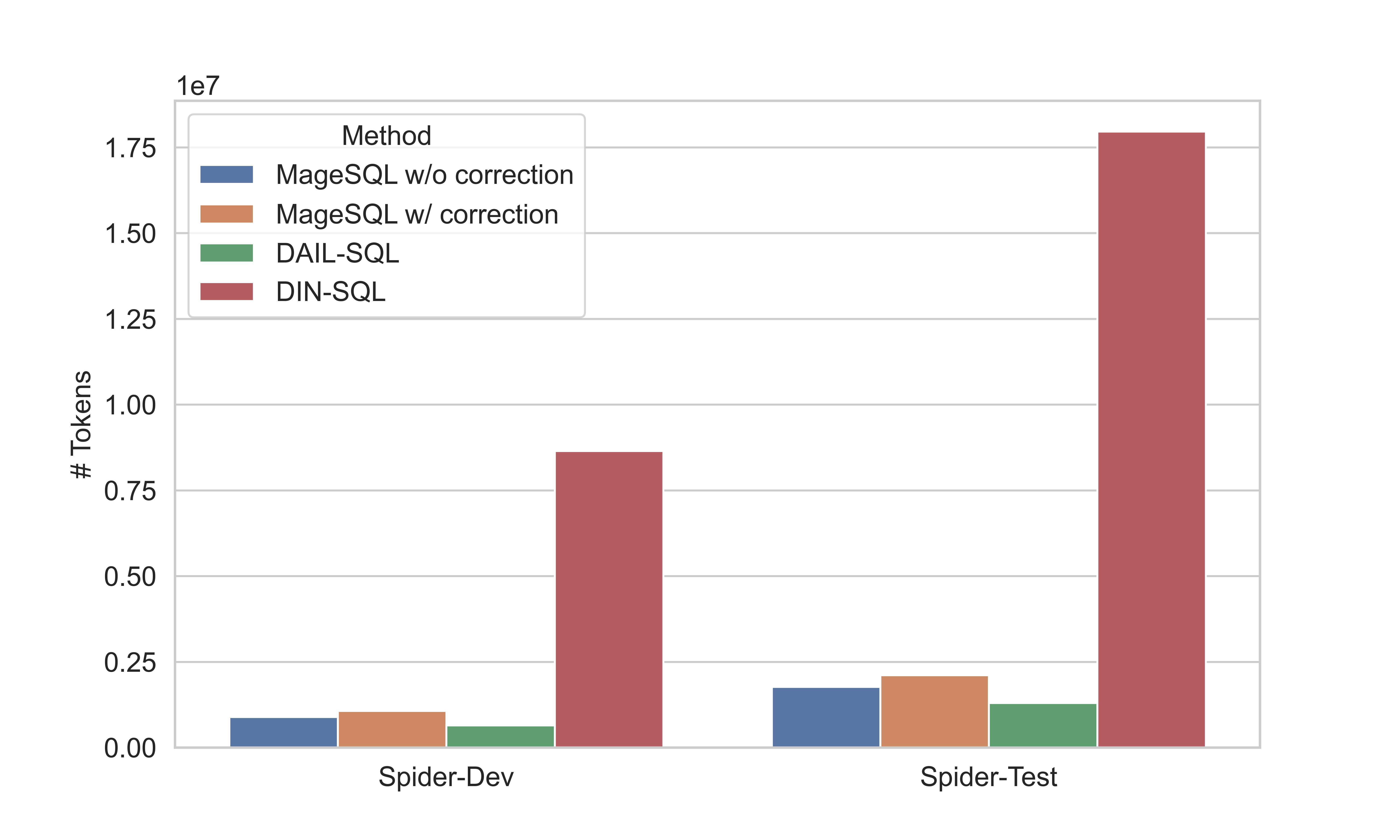}
	\caption{Cost of Different LLM-based Methods with 5-shot learning on the Spider Dataset.}
	\label{fig:cost}
\end{figure}

Finally, we report the cost of \name and other recent LLM-based methods, DIN-SQL and DAIL-SQL, in Figure~\ref{fig:cost}.
Based on the official OpenAI pricing mechanism~\footnote{https://openai.com/api/pricing/}, we use the total number of tokens in all prompts as the evaluation metric.
We can see that DIN-SQL involves the most overhead in cost since it adopts the chain-of-though method and requires 3 prompts for each instance.
Our method requires an extra prompt to generate the SQL to find examples with similar graph embeddings, though such prompts are relatively short since there is no demonstration.
Thus, the cost (even without error correction) is slightly higher than that of DAIL-SQL.
Meanwhile, error correction did not introduce much additional cost to our method.
The reason could be due to the strategy scheduling efforts shown in Section~\ref{subsec-error} that only send the challenging cases to the prompt-based error correction.

\subsection{Impact of \name Design Choices}\label{subsec-more}
\begin{figure}[ht]
	\centering
	\begin{subfigure}{\columnwidth}
		\centering
		\includegraphics[width=0.9\textwidth]{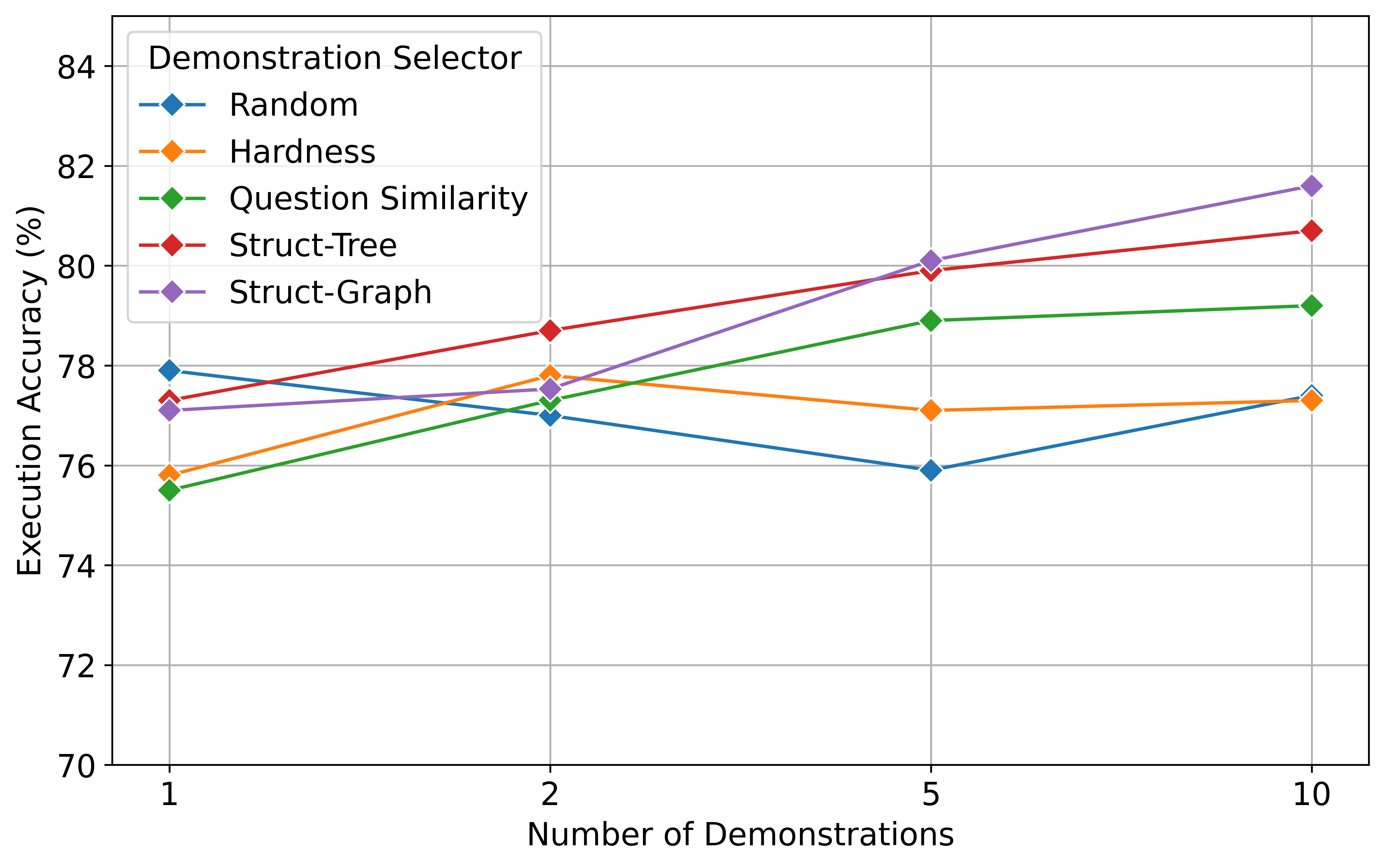}
		\caption{Spider-Dev}
	\end{subfigure}
	\begin{subfigure}{\columnwidth}
		\centering
		\includegraphics[width=0.9\textwidth]{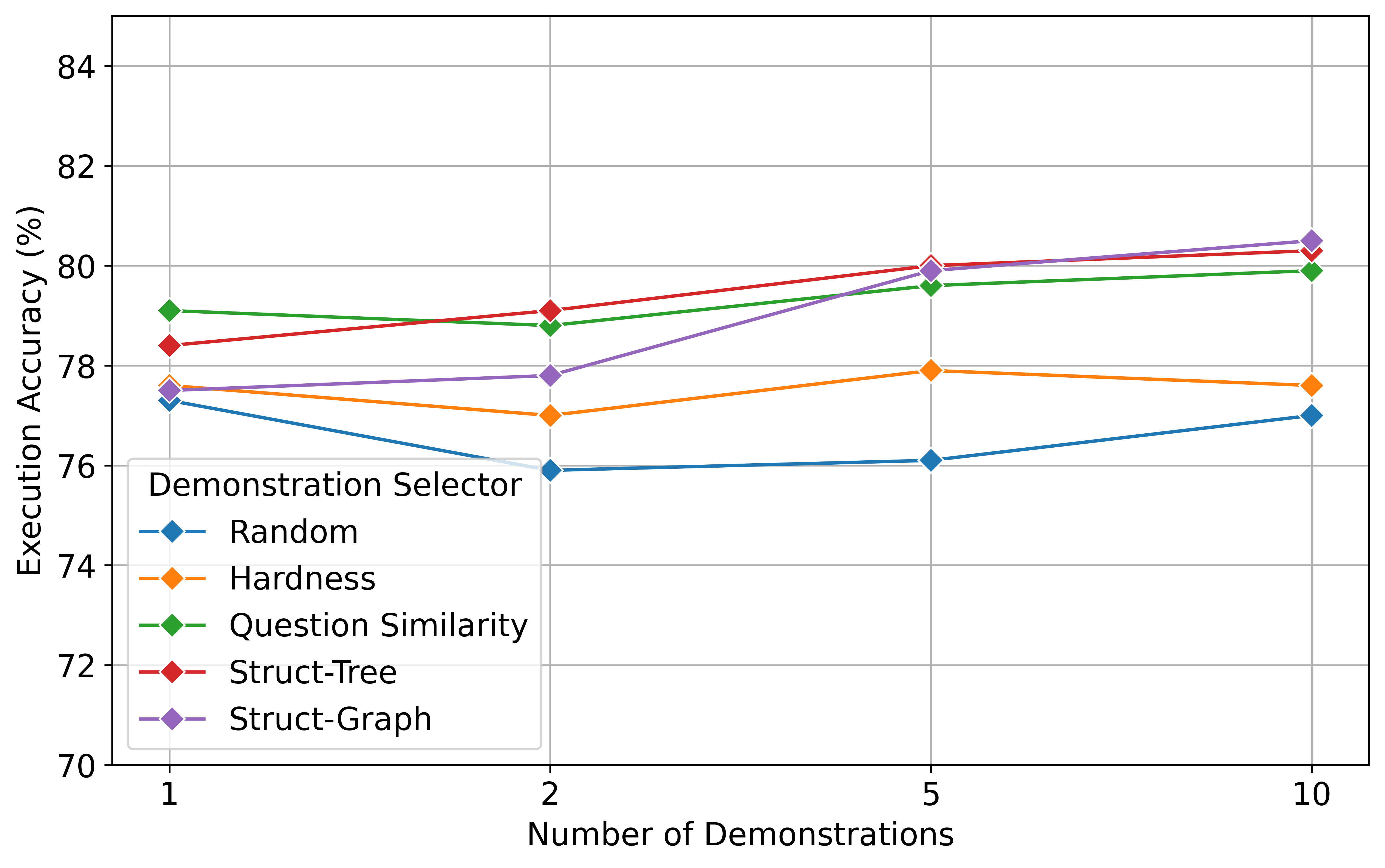}
		\caption{Spider-Test}
	\end{subfigure}
	\caption{Effect of numbers of demonstrations for Spider datasets based on GPT-3.5. The result of zero-shot learning on the dev and test set is 75.4 and 76.0, respectively.}
	\label{fig:dnum}
\end{figure}

\begin{table}[ht]
	\caption{Results on Different Demonstration Selection Strategies (EX \%).}\label{tbl-demo}
	\begin{tabular}{l|cc}
		\toprule
		Method & Spider-Dev & Spider-Test \\
		\midrule
		Zero-shot Learning & 77.6 & 77.9  \\
		Random & 81.8 & 81.9  \\
		Hardness & 83.5 & 83.1 \\
		Question Similarity & 84.0 & 84.4 \\
		Struct-Tree & 84.9 & 86.6 \\
		Struct-Graph & \textbf{87.4} & \textbf{86.8} \\
		\bottomrule
	\end{tabular}
\end{table}

Next, we conducted more experiments to analyze the effects of our proposed techniques.
We will use Execution Accuracy (EX) as the evaluation metric because it is more appropriate for LLM-based solutions.

We first look at the effect of different demonstration selection strategies in Table~\ref{tbl-demo}.
The methods Random, Hardness, Question Similarity and Struct-Tree were previously introduced in Section~\ref{subsec-demo};
while Struct-Graph is the graph embedding-based method in Section~\ref{subsec-gcl}.
Since different strategies might need different numbers of demonstration examples to achieve the best performance in few-shot learning, and such numbers don't differ much (no more than 10-shot), we report the best performance of all methods that might not have the same number of demonstrations.
Generally speaking, we observe that Random performs worst, and in fact, it is close to a zero-shot learning setting.
This clearly illustrates that bad demonstrations might harm the results in some cases.
Among all the methods, Struct-Graph achieves the best result since it could provide useful examples for some difficult instances to help LLM generate the corresponding SQL.
Although Hardness can reach similar objectives, its selection criteria are too heuristic and might not be able to find proper examples.

We then investigate the effect of a number of demonstrations.
As shown in Figure~\ref{fig:dnum}, the results of most methods tend to be better with more examples.
The exceptions are in the Random and Harness cases.
The reason could be that they both include the process of randomly selecting examples from a set of candidates and thus might not always select high-quality ones.
We also tried to include more than 10 examples as demonstrations. 
However, the results do not improve.
Therefore, we stop with the maximum number of examples as 10.

We also show the effect of error correction methods with execution accuracy (EX) as the metric.
The results are as follows: on Spider-dev, the result of execution accuracy without and with error correction is 84.5 and 87.4, respectively.
On the Spider-test, execution accuracy without and with error correction is 84.7 and 86.8, respectively.
With the help of error correction, we achieve up to 2.9\% performance gain on all datasets.
It illustrates that the error correction mechanism could help address some errors from the SQL generated from the initial prompt.
The reason could be that the pre-defined instructions could provide more useful insights for the second prompt to generate the correct query.
Besides, the rule-based approach could also help fix some instances where the semantics are correct but fail in execution just because of minor issues, e.g., different letter cases in predicate values, extra symbols like quota, etc.

\subsection{Results on the BIRD Dataset}\label{subsec-bird}

\begin{figure}[ht]
	\centering
	\includegraphics[width=0.4\textwidth]{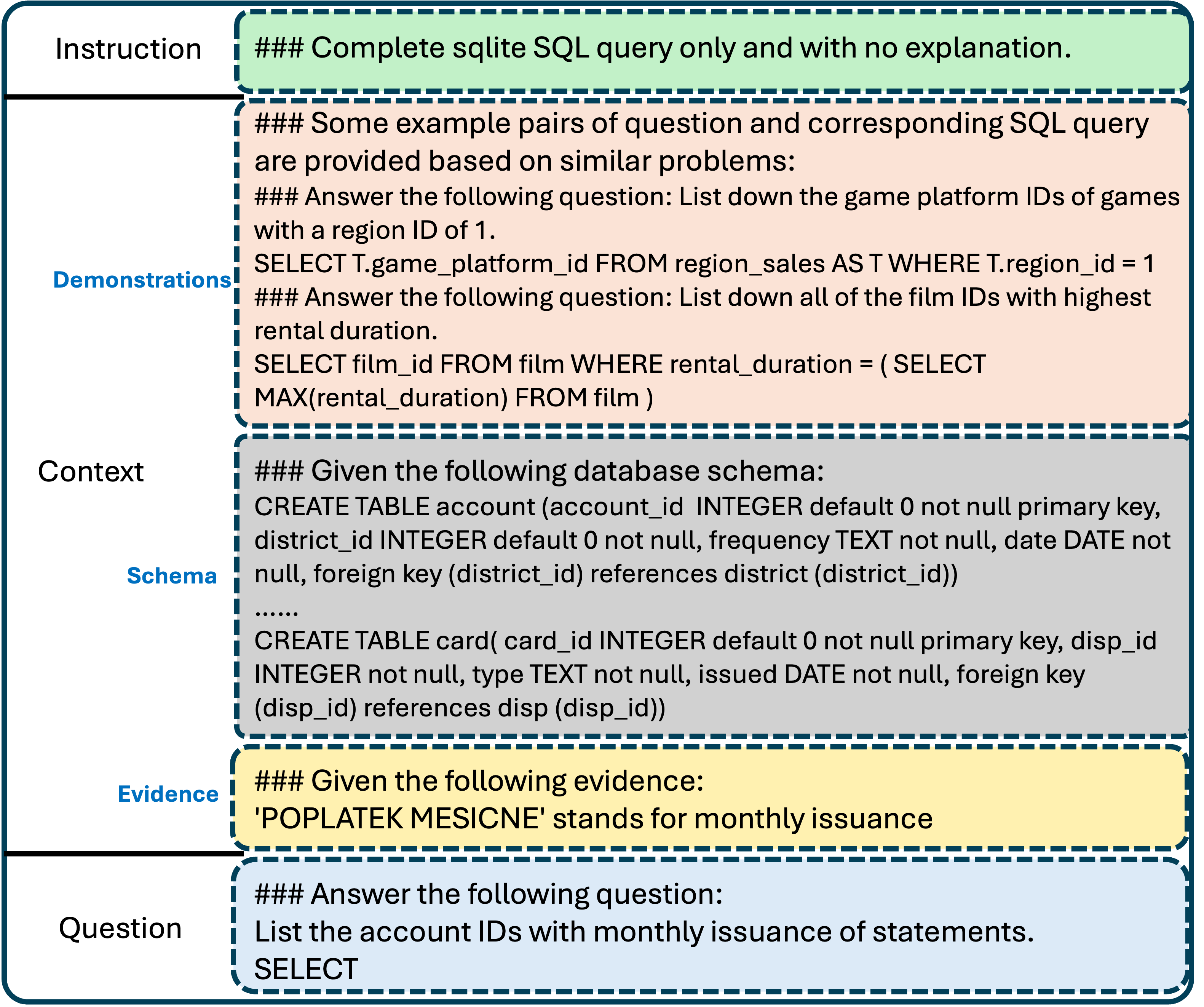}
	\caption{The Prompt Template for BIRD Dataset.}
	\label{fig:birdp}
\end{figure}

We also tested our proposed solution on the BIRD-dev dataset.
Compared with Spider, there are more tables in each database of the BIRD dataset and the model needs to accurately identify the relevant tables to answer a question. 
Meanwhile, the BIRD dataset provided additional ``evidence'' information, which are paragraphs of descriptions about databases and tables to assist disambiguation in the questions with such external knowledge.
Therefore, it is essential to include them in the prompt template to help with question understanding, as previous studies on the BIRD dataset have done.
To this end, we slightly modify the previous prompt template to satisfy the need of BIRD as shown in Figure~\ref{fig:birdp}.
Specifically, we add a section of 
"Evidence" (yellow box) at the end of Context in the prompt to accommodate such information.

\begin{table}[ht]
	\centering
	\caption{Results on the BIRD-dev dataset with performance breakdown based on difficulty levels (EX \%).}\label{tbl-bird}
	%\vspace{-2mm}
	\begin{tabular}{l|cccc}
		\toprule
		Method &  Simple & Moderate & Challenging & Overall \\
		\midrule
		RESDSQL & 53.5 & 33.3 & 16.7 & 43.9 \\
		C3 & 58.9 & 38.5 & 31.9 & 50.2 \\
		DAIL-SQL & 63.0 & 45.6 & 43.1 & 55.9 \\
		CodeS & 65.8 & \textbf{48.8} & 42.4 & 58.5 \\
		SuperSQL & 66.9 & 46.5 & 43.8 & 58.5 \\
		\name (GPT-4) & \textbf{68.54} & 48.06 & \textbf{51.03} & \textbf{60.69} \\
		\bottomrule
	\end{tabular}
\end{table}

Here we use the representative previous studies compared in the recent work SuperSQL~\cite{DBLP:journals/pvldb/LiLCLT24} as baseline methods.
The results of baseline methods are copied from~\cite{DBLP:journals/pvldb/LiLCLT24}.
For a method with multiple variants, we reported the one with the best results.
For example, RESDSQL~\cite{DBLP:conf/aaai/Li00023} is corresponding to RESDSQL-3B;
DAIL-SQL~\cite{DBLP:journals/pvldb/GaoWLSQDZ24} is the version with Self-Consistency (SC);
and CodeS~\cite{DBLP:journals/pacmmod/LiZLFZZWP0024} is corresponding to SFT CodeS-15B.
The results shown in Table~\ref{tbl-bird} illustrated that \name achieved the best overall performance among all methods.
Specifically, it outperformed the state-of-the-art method SuperSQL~\cite{DBLP:journals/pvldb/LiLCLT24} by 2.19\% in execution accuracy.
Compared with the Spider dataset, BIRD is more challenging due to its huge database volumes and much larger number of numbers in a database.
\name could alleviate such issues with the help of high-quality demonstration examples and the ability to fix minor errors in the model output.
We observe that the advantage of \name over other baseline methods is more obvious in the difficulty level of ``Challenging'' which is consistent with that on the Spider dataset.

\subsection{Error Analysis}\label{subsec-case}

We further conducted an in-depth error analysis to provide useful insights about our techniques.
To begin with, we compiled statistics on the instances in which our proposed solution failed in the Spider dataset even after error correction.
Based on the practice of previous studies~\cite{DBLP:conf/nips/PourrezaR23,DBLP:journals/corr/abs-2307-07306,DBLP:journals/pvldb/FuLWLTS23}, we made the category of errors as following:
\begin{compactitem}
	\item Syntax: There are syntax errors, and the generated SQL cannot be executed.
	\item Structure: The generated SQL failed to identify or make obvious errors in the structure of a query, such as those with multi-way join and conjunction. 
	\item Schema: The error are related to the schema information of database.
	\item Name and Semantics: The error is related to the semantics of table/attribute names or values in predicates.
	\item Aggregation: The error is related to aggregations.
\end{compactitem}

\begin{table}[ht]
	\centering
	\caption{Error Analysis on Spider Dataset (\%). The number is the percentage in all incorrect instances but not all instances in the dataset.}\label{tbl-error}
	\begin{tabular}{lcc}
		\toprule
		Error Category & Dev & Test \\
		\midrule
		Syntax &  9.6 & 10.1 \\
		Structure & 58.9 & 20.8 \\
		Schema & 45.5 & 25.3 \\
		Name and Semantics & 28.2 & 43.8 \\
		Aggregation & 39.7 & 18.1 \\
		\bottomrule
	\end{tabular}
\end{table}

The results of statistics are shown in Table~\ref{tbl-error}.
We recognize that one incorrect instance could involve multiple types of errors.
Therefore, the overall number could exceed 100\% in each dataset.
We can see that most errors in the Dev set come from the Structure and Schema categories, which correspond to the instances in the hard and extra categories. 
At the same time, the challenges in the Test set mainly come from Name and Semantics, where many cases require the LLM to understand not only the question but also the semantics of table and attribute names.
In such cases, errors in the ``Name and Semantics'' category always happened together with those in the ``Schema'' one.

\begin{figure*}[ht]
        \begin{subfigure}{\columnwidth}
		\centering
		\includegraphics[width=0.8\textwidth]{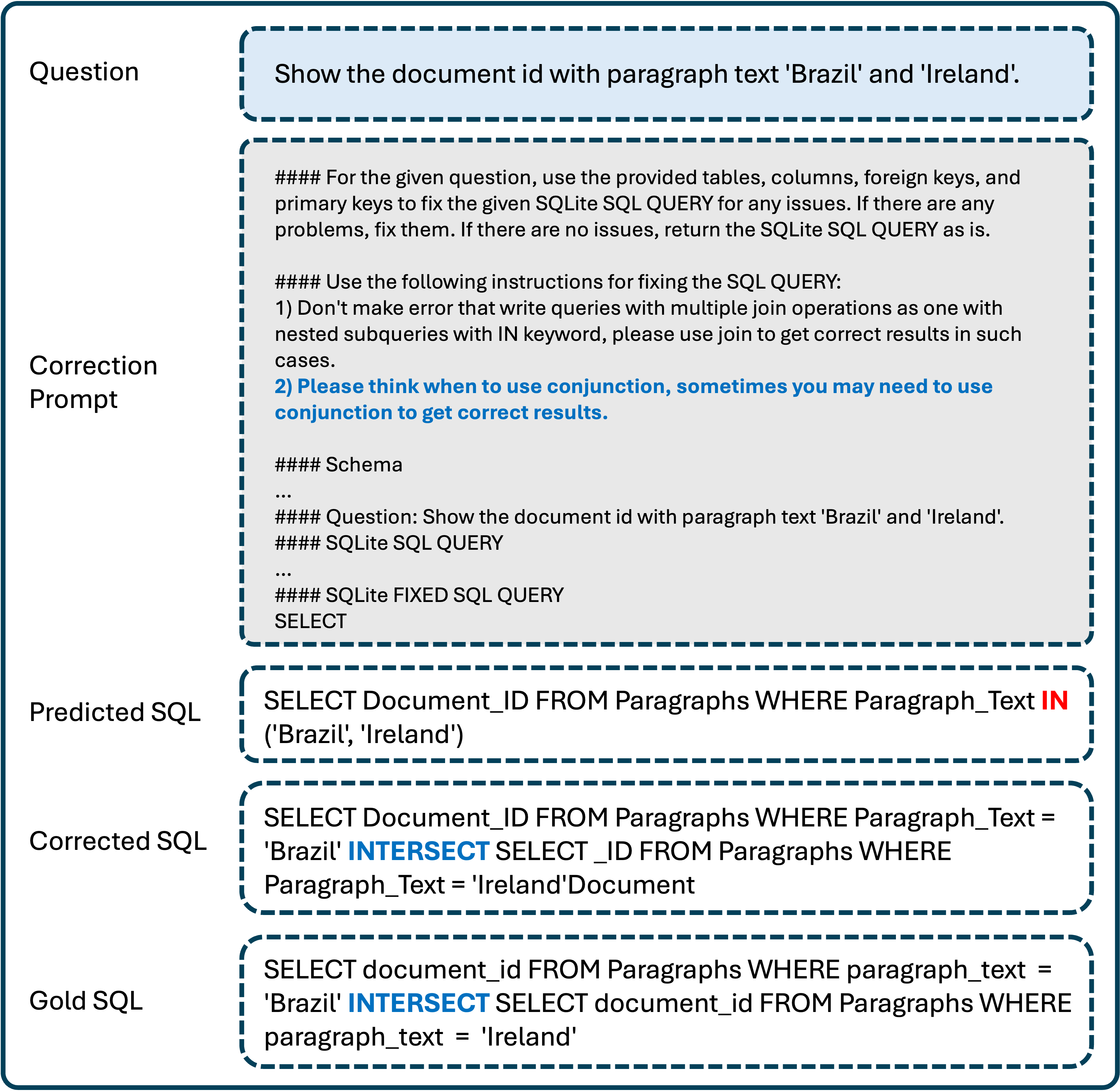}
		\caption{Successful example of the prompt-based correction method}
		\label{fig:dnumd}
	\end{subfigure}
	\begin{subfigure}{\columnwidth}
		\centering
		\includegraphics[width=0.8\textwidth]{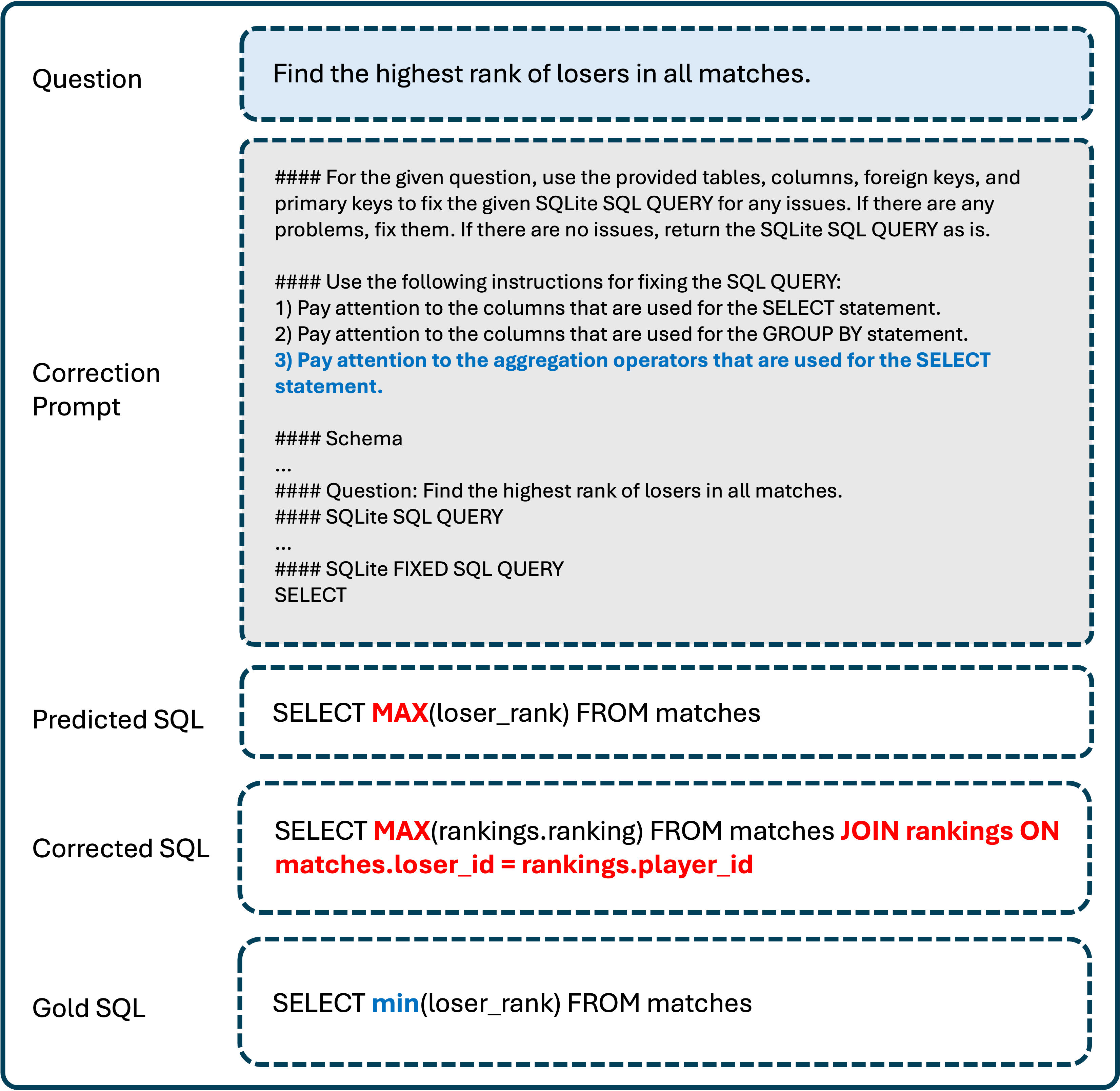}
		\caption{Failure example of the prompt-based correction}
		\label{fig:dnumt}
	\end{subfigure}
	\centering
	\caption{Examples of Error Correction with Prompt.}
	\label{fig:errorcase}
\end{figure*}

Next, we conduct a case study about our prompt-based error correction method in Figure~\ref{fig:errorcase}.
The detailed schema and query in correction prompts are omitted due to space limitations.
Figure~\ref{fig:errorcase}(a) illustrates a scenario where the prompt-based method can successfully identify and fix the error.
We can see that the initial output of LLM has errors in identifying the need to use conjunctions.
We apply a prompt with the template shown in Figure~\ref{fig:error} and utilize the guidelines ``Please think when to use a conjunction; sometimes you may need to use a conjunction to get correct results".
Then, such errors could be fixed with another round of prompts.
At the same time, the prompt-based strategy may also fail in some instances, as illustrated in Figure~\ref{fig:errorcase}(b). 
In this example, the generated SQL incorrectly interprets the semantics of ``highest rank''. 
Despite adhering to the guidelines that emphasize appropriate aggregation operators in the SELECT statement, the final output is still an erroneous SQL statement. 
This error might stem from the LLM's limitations in handling the semantic nuances of the query.
We are thrilled to continue exploring how to fix such issues via prompt or advanced rules in future work.

\section{Related Work}\label{sec-related}

% \subsection{Solutions for Text-to-SQL}\label{subsec-relt2s}

The Text-to-SQL problem has been well-explored for many years.
Earlier studies~\cite{DBLP:journals/pvldb/KimSHL20,DBLP:journals/pvldb/SahaFSMMO16} first parsed the natural language question into intermediate results and then developed different kinds of rules to map it into the abstract syntax tree of SQL so as to generate the final query.
The limitation of such methods is that they always perform poorly when adapted to new domains.
To address such issues, another category of studies employed deep learning techniques, converting the Text-to-SQL problem into a machine translation task and train deep learning models to generate SQL statements.
Based on the different ways of decoding, the solutions can be categorized into sequence~\cite{DBLP:conf/emnlp/ScholakSB21,DBLP:conf/acl/HuZJLZCLPWHZGDL23}, tree~\cite{DBLP:conf/naacl/RubinB21,DBLP:conf/emnlp/QiTHW0ZWZL22} and graph~\cite{DBLP:conf/acl/BoginBG19,DBLP:conf/acl/CaoC0ZZ020,DBLP:conf/aaai/LiHCQ0HHDSL23} based ones.
With the advances of PLM, recent studies developed the solutions by pre-training a language model for structured data to support various tasks including text-to-SQL, such as TAPEX~\cite{DBLP:conf/iclr/LiuCGZLCL22} and GraPPa~\cite{DBLP:conf/iclr/0009WLWTYRSX21}.
Meanwhile, another category of studies first developed a sketch template of SQL and then use decoder based models to fill the empty slots in the template~\cite{DBLP:journals/coling/ChoiSKS21,DBLP:conf/emnlp/GanCXPWDZ21,DBLP:journals/pvldb/FuLWLTS23}.
This approach could avoid generating invalid SQL queries while fully utilizing the powerful decoders.

Recent efforts mainly aimed at leveraging the power of LLMs to understand the natural language question and generate the SQL accordingly.
Due to their profound capabilities, LLM-based solutions have significantly outperformed previous methods and achieved state-of-the-art performance.
C3~\cite{DBLP:journals/corr/abs-2307-07306} proposed a precise prompt instruction for the zero-shot setting.
Rajkumar et al.~\cite{DBLP:journals/corr/abs-2204-00498} made an investigation of prompt strategies over different kinds of LLMs.
Nan et al.~\cite{DBLP:conf/emnlp/Nan0ZRTZCR23} explored different prompt templates for both zero-shot and few-shot settings.
DIN-SQL~\cite{DBLP:conf/nips/PourrezaR23} employed the chain-of-thought reasoning strategy and divided the text-to-SQL problem into 3 stages to solve issues from different aspects in each stage respectively.
DAIL-SQL~\cite{DBLP:journals/pvldb/GaoWLSQDZ24} conducted an empirical study on the different combinations of prompt instructions and demonstration selection strategies, which overlaps with our work but focuses on different aspects of the problem.
SuperSQL~\cite{DBLP:journals/pvldb/LiLCLT24} explored the combination of different components in prompt templates and summarized optimal solutions for different tasks.
CodeS~\cite{DBLP:journals/pacmmod/LiZLFZZWP0024} aims at fine tuning smaller language models instead of constructing prompts, which has a different objective.

Another line of studies lie in the aspect of benchmarking the text-to-SQL tasks.
There are two categories of datasets: general purposed and domain specific ones.
The general purposed works aimed at building cross-domain datasets that have a broad coverage.
Examples include WikiSQL~\cite{DBLP:journals/corr/abs-1709-00103}, Spider~\cite{DBLP:conf/emnlp/YuZYYWLMLYRZR18}, KaggleDBQA~\cite{DBLP:conf/acl/LeePR20} and BIRD~\cite{DBLP:conf/nips/LiHQYLLWQGHZ0LC23};
While the domain specific works focused on developing datasets for a specific application domain, such as Yelp and IMDB~\cite{DBLP:journals/pacmpl/Yaghmazadeh0DD17}, FINSQL~\cite{DBLP:conf/sigmod/ZhangMFMG0LL24} and BookSQL~\cite{DBLP:conf/naacl/KumarDHSM24}.
In this work, we focused on developing general solution for the text-to-SQL problem and thus conducted evaluation over the general purposed benchmarking datasets.

% \subsection{Large Language Models}\label{subsec-relllm}
%To take advantage of LLMs, a large amount of research efforts have been made in investigating the learning ability of LLMs~\cite{DBLP:conf/acl/GaoFC20,DBLP:conf/nips/PerezKC21,DBLP:conf/iclr/WeiBZGYLDDL22}, developing reasoning strategies~\cite{DBLP:conf/nips/Wei0SBIXCLZ22,DBLP:conf/iclr/YaoZYDSN023,DBLP:conf/iclr/KhotTFF0CS23} as well as ensuring the consistency of the results~\cite{DBLP:conf/emnlp/ZhouHMBN22,DBLP:conf/iclr/0002WSLCNCZ23}.
% Another line of research works lies in improving the usability of LLM~\cite{DBLP:conf/nips/SchickDDRLHZCS23,DBLP:journals/corr/abs-2308-08155}, especially supporting complicated applications that require the cooperation of multiple agents.

\section{Conclusion}\label{sec-conc}

In this paper, we conducted a systematic study of the Text-to-SQL problem.
We proposed \name, a novel framework that aimed at improving the prompt engineering process over LLM so as to help generate high-quality SQL statements as the solution.
To this end, we proposed technical contributions from two aspects: (1) develop two effective demonstration selection strategies based on the structure and semantics of SQL queries to improve the in-context learning process; (2) propose an error correction module that could find and correct the potential errors in the output of LLMs.
Experimental results on public benchmarking datasets showed that our proposed methods could obviously improve the overall performance compared with previous solutions.
% For future work, we aim to integrate complicated reasoning techniques with our proposed framework to further improve performance and extend our framework to queries over data with different modalities, such as JSON, graph, and image.

\balance
\bibliographystyle{IEEEtran}
\bibliography{ref/sql,ref/llm,ref/other}

\end{document}